\documentclass[12pt,english]{article}
\usepackage[latin9]{inputenc}
\usepackage{geometry}
\usepackage{babel}
\usepackage{float}
\usepackage{amsmath}
\usepackage{amsthm}
\usepackage{amssymb}
\usepackage{graphicx}
\usepackage{setspace}
\usepackage[unicode=true]
 {hyperref}

\makeatletter

\providecommand{\tabularnewline}{\\}

\theoremstyle{plain}
\newtheorem{thm}{\protect\theoremname}

\usepackage{amsthm}
\usepackage{setspace}
\usepackage{sectsty}
\usepackage{anysize}
\usepackage{datetime}
\usepackage{pdflscape}
\usepackage{times}
\usepackage[small]{caption}
\usepackage[bottom,hang,flushmargin]{footmisc}
\usepackage{subfig}
\@ifundefined{showcaptionsetup}{}{%
 \PassOptionsToPackage{caption=false}{subfig}}
\usepackage{subfig}
\makeatother

\providecommand{\theoremname}{Theorem}

\begin{document}
\title{Taxing dissent: The impact of a social media tax in Uganda}
\author{\noindent Levi Boxell,\textbf{ }\textit{Stanford University}\textbf{}\thanks{E-mail:\ lboxell@stanford.edu and zst@luskin.ucla.edu. We thank Nicholas
Bloom, Matthew Gentzkow, Guy Grossman, William Hobbs, Matthew O. Jackson,
Jenn Larson, Janet Lewis, Vianney Mbonigaba, Nicholas Obradovich,
Chiara Pasquini, Margaret Roberts, Jesse M. Shapiro, and Chuan Yu
for their suggestions and comments, along with participants at Stanford
University. Funding was provided by the National Science Foundation
(grant number: DGE-1656518) and the Institute for Humane Studies.}\textbf{}\\
Zachary Steinert-Threlkeld,\textbf{ }\textit{UCLA}}
\date{\monthname\ \number\year}
\maketitle
\begin{abstract}
We examine the impact of a new tool for suppressing the expression
of dissent\textemdash a daily tax on social media use. Using a synthetic
control framework, we estimate that the tax reduced the number of
georeferenced Twitter users in Uganda by 13 percent. The estimated
treatment effects are larger for poorer and less frequent users. Despite
the overall decline in Twitter use, tweets referencing collective
action increased by 31 percent and observed protests increased by
47 percent. These results suggest that taxing social media use may
not be an effective tool for reducing political dissent.

\bigskip{}
\end{abstract}
\pagebreak{}

\begin{spacing}{1.4}

\section{Introduction}

Autocratic regimes have become increasingly interested in digital
censorship, and new strategies for controlling online information
are rapidly spreading. After tests of internet shutdowns in the Xinjiang
province of China in 2009, governments shut down internet access nearly
190 times in 2017 (Griffiths 2019). China has also been actively exporting
its digital repression strategies to other countries. In 2017, officials
from Uganda met with members of a Chinese government subsidiary to
build capacity to monitor and prevent social media abuse (Mugerwa
2017). Later, Uganda innovated on the set of tools available to authoritarian
regimes by implementing a 200 shilling daily tax on users of social
media.\footnote{The tax is equivalent to \$1.60 per day for the United States (normalized
by GDP per capita PPP) and applies to over 50 social media applications
including Facebook, WhatsApp, and Twitter.} Critics view the tax as an attempt to suppress political dissent
ahead of upcoming elections, and even the Ugandan President has stated
that reducing ``gossip'' was an intended outcome of the tax (Crabtree
2018).

In this paper, we examine the impact of the social media tax on social
media use and levels of collective action in Uganda. Using georeferenced
tweets from across sub-Saharan Africa and exploiting the unexpected
implementation of the social media tax, we compare outcomes in Uganda
relative to a synthetic control composed of non-treated countries
with similar pre-treatment trends (Abadie et al. 2010).

There was a large and sustained decrease in the number of active users
on Twitter after the tax. The tax decreased the number of unique Twitter
users in each period by 13 percent on average. New account registrations
also dropped by 20 percent. Poorer and less frequent users exhibited
a larger change in behavior, with a 17 percent decrease in non-Apple
users (a proxy for wealth) and an 18 percent decrease in users who
tweet less than once per day. These estimated effects are large relative
to a placebo distribution of non-treated countries.

Despite the reduction in overall social media use, tweets referencing
collective action increased by 31 percent in the period following
the implementation of the tax. In contrast, there was no meaningful
increase in the overall level of tweets or tweets using political
phrases not related to collective action.\footnote{Increasing the availability of information on collective action is
particularly concerning for authoritarian regimes. While the impact
of increasing the flow of information on grievances against the government
is theoretically ambiguous depending on citizen expectations, information
on the events themselves nearly always increases the likelihood of
successful coordination (Little 2016). Ananyev et al. (2017) make
a related argument that less competent governments will be more likely
to invest in coordination censorship than content censorship. Observed
behavior of Chinese censorship corroborates the focus on collective
action information (King et al. 2013).} The period after the tax's implementation also experienced a 17 percent
increase in the number of unique users discussing collective action.
This increase in discussions of collective action online is mirrored
by increases in physical protests and riots. There was a 47 percent
increase in observed protests and riots in the subsequent period in
Uganda relative to a synthetic control. Again, these estimated effects
are large relative to the placebo distribution of non-treated countries.

The tax could increase collective action through two potential mechanisms.

First, it could provide a signal of the regime's type. The high salience
of the tax as a form of digital repression may have decreased coordination
costs for collective action by creating a focal point or providing
common information (Chwe 2000; Tucker 2007; Egorov et al. 2009; Chwe
2013). Alternatively, the signal from the tax could have shifted the
perceived returns to succesful protest and regime change. Qualitative
evidence from news coverage of the tax and the language used by oppositional
leaders on Twitter corroborate the tax's role as a rallying point
for collective action, and previous work has also suggested that the
salience of repression is an important determinant of the overall
effect on dissent (Roberts 2018).

Second, the tax could increase collective action by altering the composition
or structure of the network. In network models of collective action,
an individual's decision to participate often depends on the extent
to which users within her \emph{local} network participate (Chwe 2000;
Watts 2002; Barbera and Jackson 2019).\footnote{Local network participation can be important for inference about global
participation, information on the event itself, and consumption value
from social network participation. The importance of local network
participation in determining individual turnout for protests is also
highlighted in recent empirical work (Burzstyn et al. 2019).} If the likelihood of participating in collective action is positively
correlated with willingness to pay for Twitter, then the tax would
increase the density of activists on Twitter, increasing the likelihood
of collective action.\footnote{See the Online Appendix for a model highlighting this relationship.
Previous work has also pointed out that the relationship between group
size and collective action varies depending on other forms of network
structure (Oliver and Marwell 1988; Gould 1993; Siegel 2009). Hassanpour
(2014) makes a related argument that full connectivity can hinder
collective action in threshold models. And, in models of information
overload, a small tax on sending messages will filter out low-value
messages\textemdash potentially increasing consumption of information
on collective action (Van Zandt 2004; Anderson and de Palma 2009;
Iyer and Katona 2016).} We find some empirical support for this positive correlation, as
users who are less sensitive to the tax (users with Apple devices)
are also more likely to tweet about collective action in the pre-intervention
period.

Among autocratic regimes, China's digital censorship and repression
is most studied. It has developed a sophisticated regime that focuses
on censoring information on collective action and actively produces
information to distract from sensitive topics (King et al. 2013, 2014,
2017). China also allows some forms of dissent on social media to
assist in monitoring public opinion and the performance of local officials
(Qin et al. 2017). Long-run exposure to censorship in China has reduced
the incentives to acquire uncensored information (Chen and Yang 2017).
However, the sudden and visible censorship of Instagram in China caused
citizens to actively seek out sensitive information (Hobbs and Roberts
2018).

In other contexts, disruptions to communication networks correlate
with increases in realized dissent (Hassanpour 2014), though they
also appear to weaken the ability of organized opposition groups to
coordinate (Ghodes 2015). Physical repression against dissenting elites
in Saudia Arabia reduced criticism by those directly involved, but
did not reduce overall levels of criticism (Pan and Siegel 2018).
More generally, the growth of electronic and online social networks
tends to increase protest activity (Manacorda and Tesei 2016; Enikopolov
et al. 2017; Enikopolov et al. 2018). We show that taxing social media
can effectively reduce social media use, but its high salience as
digital repression and its impact on the underlying network structure
can increase the likelihood of collective action. An important caveat
is that we study the short-run effect the tax, and the long-run effects
may be different.

\section{Ugandan Context}

President Yoweri Museveni has ruled Uganda since 1986. In this span,
the elimination of several important constitutional limits on executive
rule have allowed the 74-year-old to continue to rule. In 2005, the
two-term limit on the presidential office was removed. In 2017, the
age limit on the presidential office (75 years) was lifted\textemdash leading
to a ``brawl'' in the Ugandan parliament (BBC 2017). Museveni, who
largely gathers electoral support from rural and aging populations,
has faced increasing difficulty connecting with the youth of Uganda.
In contrast, Robert Kyagulanyi Ssentamu, a Ugandan pop-star with the
stage name Bobi Wine, has quickly risen to become the face of the
opposition party after becoming a member of parliament in 2017 (Kagumire
2018). Kyagulanyi frequently uses social media, in addition to his
music, to speak directly about government failures and the hopes of
the Ugandan people.

Within this growing discontent and opposition, Museveni and his government
actively seek to control the growth of digital dissent in Uganda.
They have arrested dissenting voices on social media for violating
the 2011 Computer Misuse Act and shut down social media services on
Election Day in 2016 (Mugerwa 2017). In 2017, Ugandan officials met
with a Chinese government-owned subsidiary to provide Uganda with
``technical capacity to monitor and prevent social media abuse''
across the more than 13 million internet users (Mugerwa 2017). Shortly
after, the Ugandan Communication Commission announced that they would
launch their own versions of popular social media platforms, mirroring
a prominent Chinese strategy for monitoring and controlling information
flows (Observer 2017). These alternative platforms have yet to materialize.

In May 2018, the Ugandan parliament passed a bill creating a 200 shilling
daily tax on the use of any of more than 50 social media applications,
along with a tax on mobile money transfers. The tax was implemented
on July 1, 2018, and is collected by telecommunication companies via
mobile money transfers (Crabtree 2018). The Online Appendix (Figure
\ref{fig:synthetic_control_effects_early_check}) shows that there
was a large spike in news articles referencing the ``social media
tax'' after July 1, 2018 and limited discussion prior to this. In
relative terms, the 200 shilling (roughly \$0.05 USD) per day tax
is large. The GDP per capita (PPP) for Uganda was \$1,864 in 2017
relative to \$59,532 in the United States, per the World Bank. Annualizing
the \$0.05 per day gives a tax burden of the social media tax at roughly
1 percent of Uganda's GDP per capita. In comparison, the median willingness-to-pay
estimate from Sunstein (Forthcoming) for Twitter is \$5.00 per month
or 0.1 percent of the United States' GDP per capita.

Many view the tax as a form of digital repression by Museveni who
hopes to limit social media use for the upcoming 2021 presidential
election (Crabtree 2018). Protests, led by Kyagulanyi, occurred shortly
after the tax's implementing, and Kyagulanyi, who announced he will
also campaign in the 2021 presidential election, has been subsequently
arrested multiple times (Al Jazeera 2019).

\section{Data and Empirical Framework}

Our primary data consists of georeferenced tweets from 50 sub-Saharan
Africa countries and territories (hereafter, ``countries'') collected
from the one percent Twitter Streaming API. For each tweet, the API
returns the text, the time of the tweet, and various user characteristics
at the time of the tweet. We pass a global bounding box to the streaming
API, delivering an estimated 1/3 to 1/2 of all georeferenced tweets
(Leetaru et al. 2013; Steinert-Threlkeld 2018). The tweets are assigned
to their respective countries using the provided geolabelling from
Twitter. We collapse observations into ten day periods and exclude
countries in the bottom quintile of unique users (e.g., Central African
Republic and Eritrea).

For data on collective action events, we merge data on riots and protests
as recorded by ACLED and ICEWS (Raleigh et al. 2010; Boschee et al.
2015). Separately, for each dataset, we compute the number of events
occuring in a given time period. We then take the average across both
datasets for a given country in a given period. We restrict attention
to countries that appear in both our ACLED and ICEWS samples.

To estimate the impact of the social media tax, we adopt a synthetic
control approach (Abadie et al. 2010). For inference, we compare the
estimated treatment effect for Uganda $\hat{\tau}_{0t}$ to the placebo
distribution of estimated effects $\mathcal{P}_{t}$ created by repeating
the synthetic control procedure for each control unit and scaling
these placebo effects by the ratio of the root mean squared prediction
error (RMSE) in the pre-intervention period for Uganda and the RMSE
for the corresponding control unit.

See the Online Appendix and the replication code for additional details.

\section{Results}

\subsection{Effect of the Tax on Social Media Use}

\begin{figure}[H]
\caption{Impact of Tax on Twitter Use\label{fig:main_effect}}
\medskip{}

\medskip{}

\subfloat[\emph{\quad{}\quad{}Panel A: Uganda vs. Select Other Countries}]{
\begin{centering}
\textit{\small{}\includegraphics[scale=0.5]{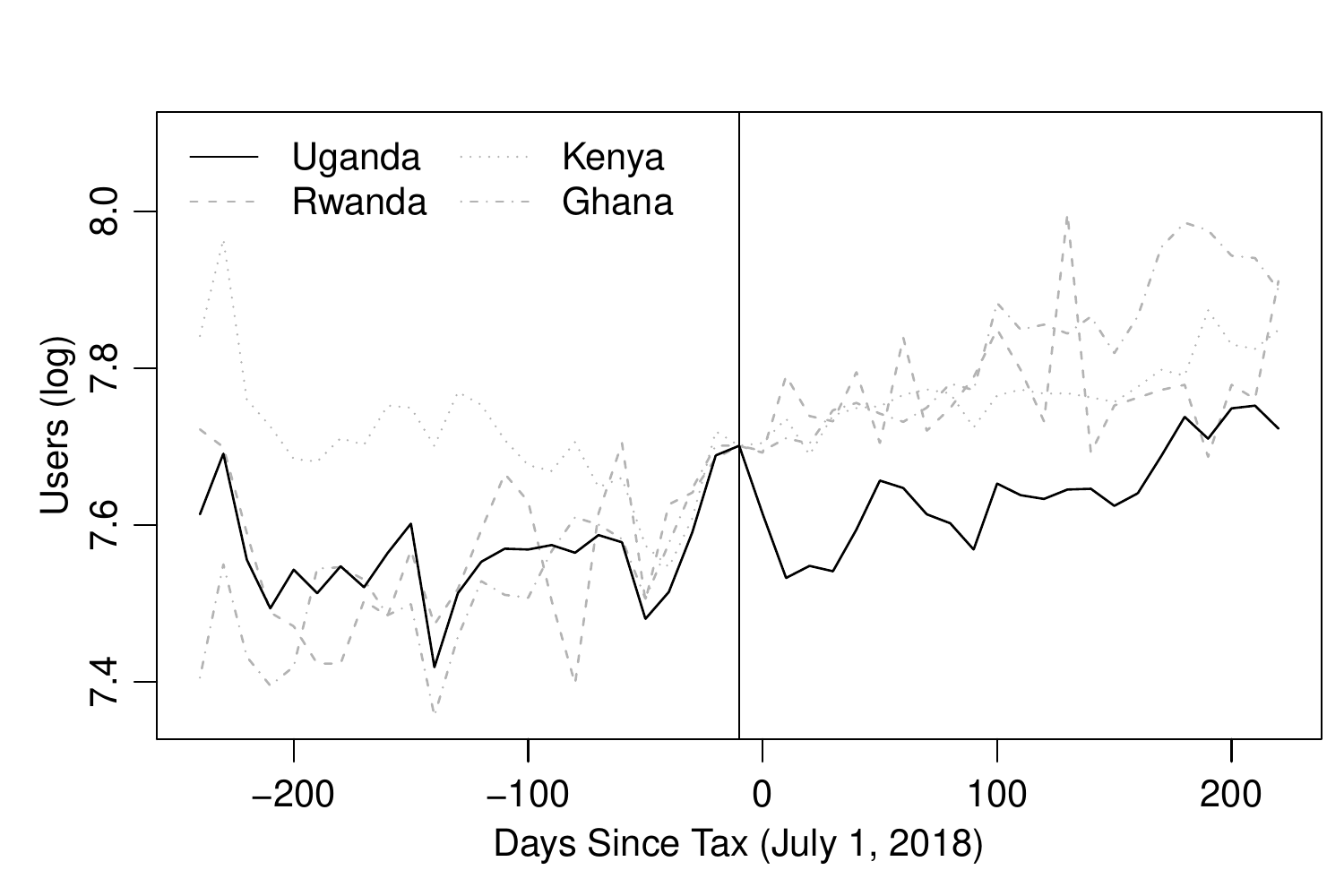}}{\small\par}
\par\end{centering}
}\subfloat[\emph{\quad{}\quad{}Panel B: Uganda vs. Average of Other Countries}]{
\centering{}\textit{\small{}\includegraphics[scale=0.5]{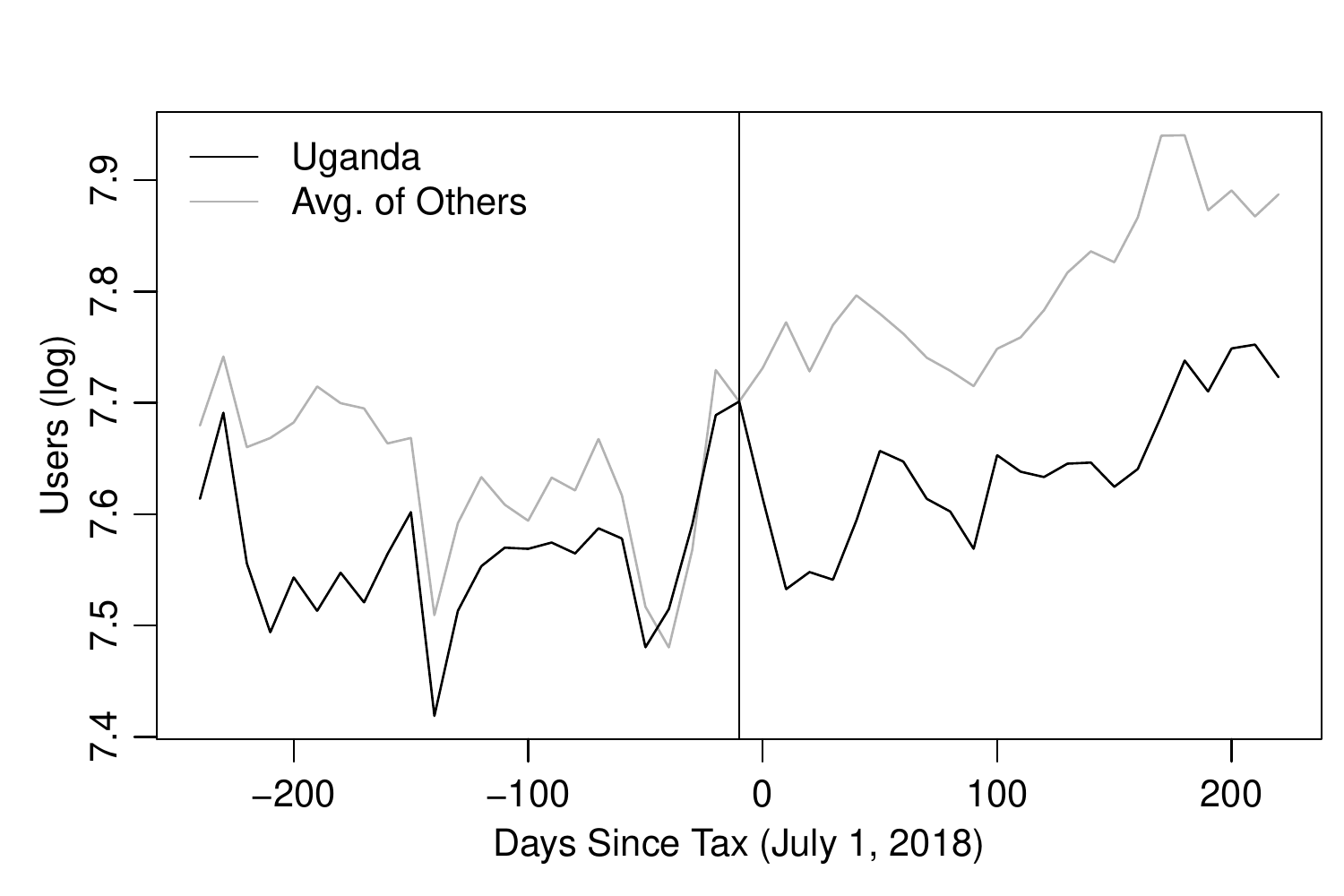}}{\small\par}}

\medskip{}
\medskip{}

\subfloat[\emph{\quad{}\quad{}Panel C: Uganda vs. Synthetic Control}]{
\centering{}\emph{\enskip{}}\textit{\small{}\includegraphics[scale=0.5]{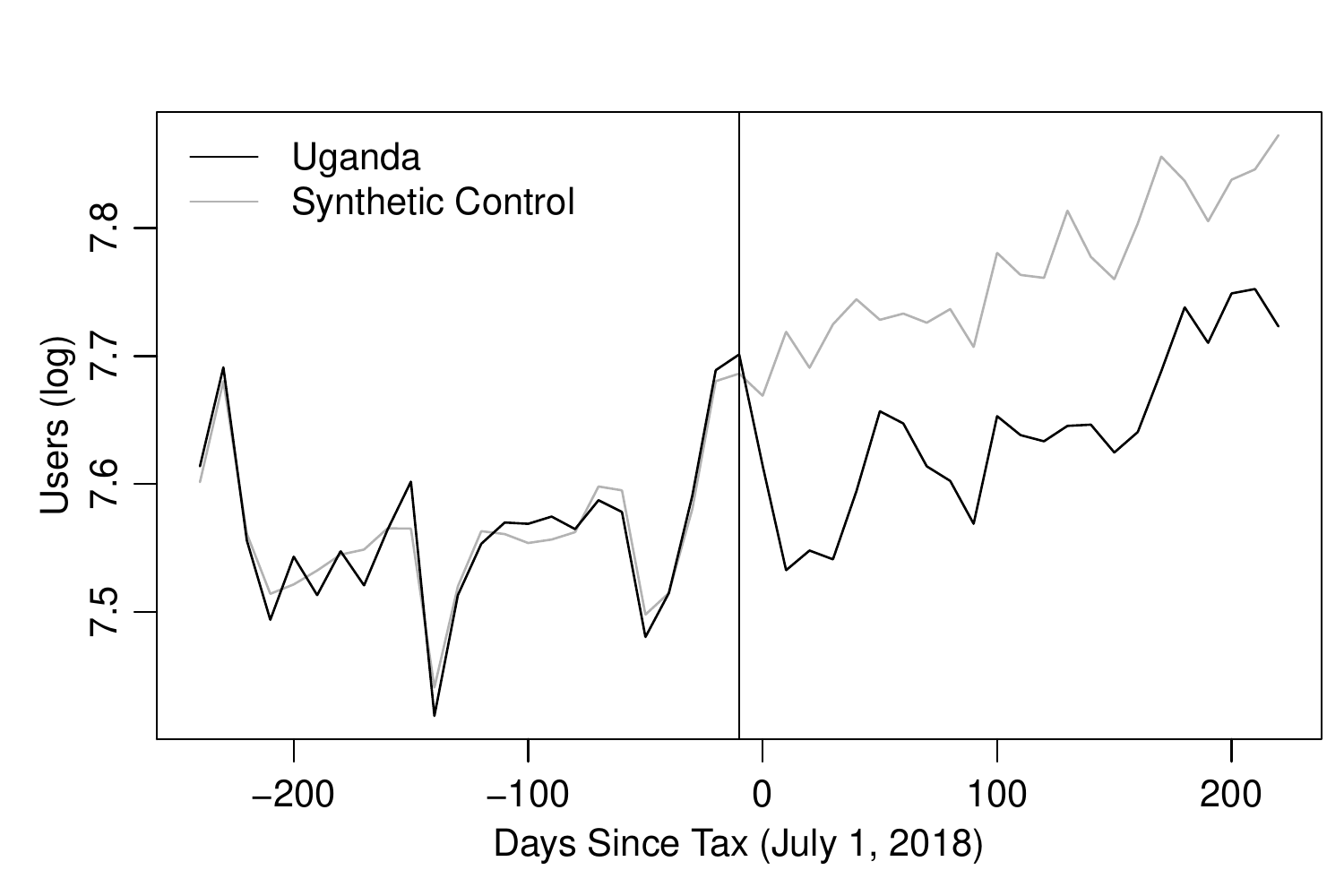}}{\small\par}}\subfloat[\emph{\quad{}\quad{}Panel D: Treatment Effect Estimates}]{
\centering{}\emph{\enskip{}}\textit{\small{}\includegraphics[scale=0.5]{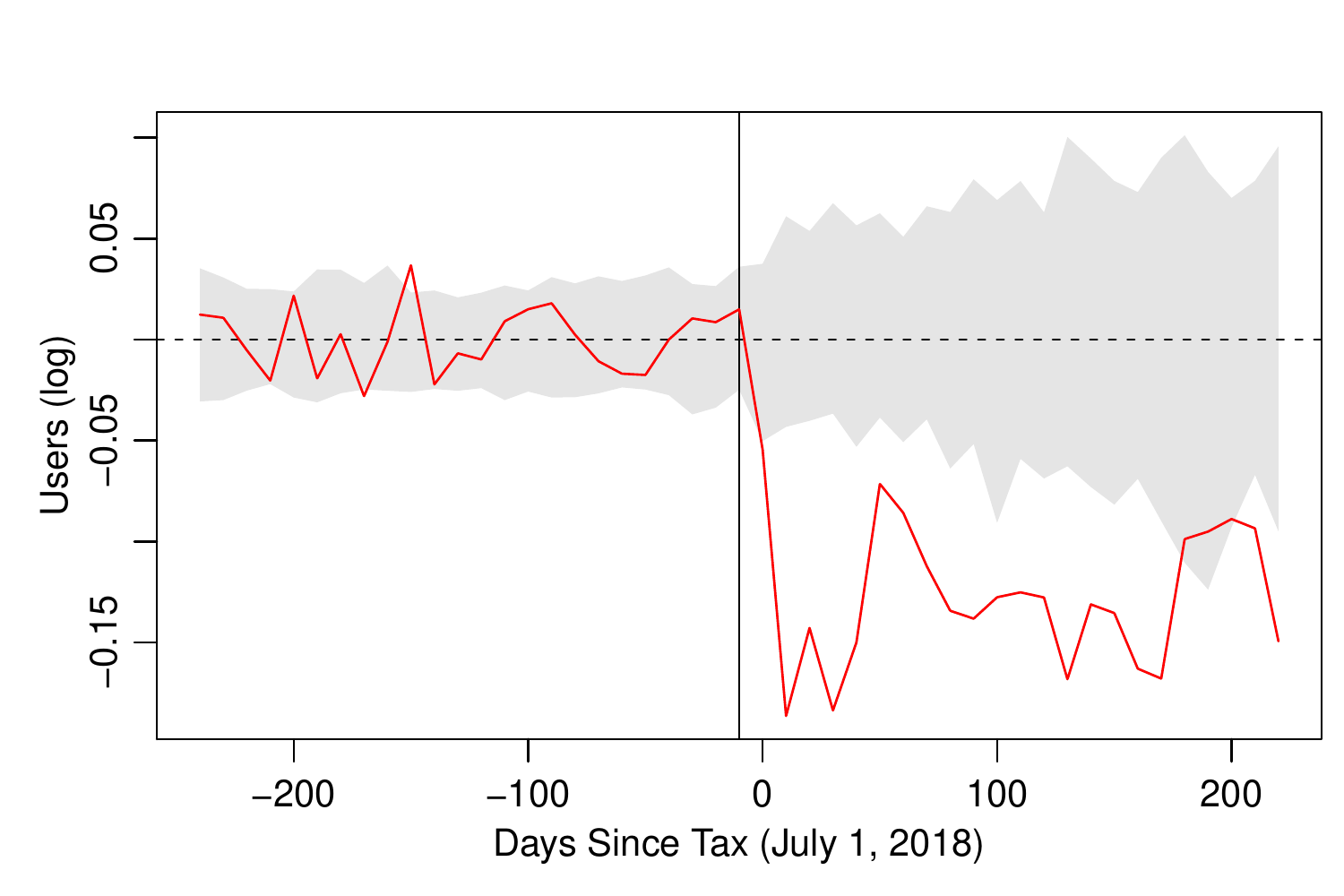}}{\small\par}}

\medskip{}

{\footnotesize{}Notes: Panel A compares the log of one plus the number
of unique, active users in each ten day period for Uganda (black)
and select other African countries (grey). Panel B compares the log
of one plus the number of unique, active users in each ten day period
for Uganda (black) and the average of the African countries and territories
kept after the sample restrictions (grey). The difference between
the log of one plus the number of users in Uganda versus the other
African countries in the period immediately prior to the intervention
($t=-1$) is subtracted from the series for each individual country
(Panel A) and the average (Panel B). Panel C plots the log of one
plus the number of unique, active users in Uganda during each ten
day period alongside the synthetic control. Panel D plots the estimated
treatment effect $\hat{\tau}_{0t}$ of the social media tax on the
log of one plus the number of unique, active users in Uganda. The
shaded region indicates the .025 and .975 quantiles of the scaled
placebo distribution $\mathcal{P}_{t}$. The vertical line indicates
the 10 day period immediately prior to when the social media tax was
implemented in Uganda (July 1, 2018).}{\footnotesize\par}
\end{figure}

Panel A of Figure \ref{fig:main_effect} plots the log of unique,
active Twitters users in Uganda relative to Rwanda, Kenya, and Ghana
in each ten day period surrounding the implementation of the social
media tax.\footnote{We define an active user to be one that has an observed tweet during
a given period. The countries selected in Panel A have the highest
weights in the synthetic control estimates.} Panel B uses the average of the other countries in the estimation
sample. In the 100 days prior to the tax, Uganda followed a similar
trend relative to the other countries. After the tax, however, Twitter
use in Uganda is systematically lower. When comparing Uganda to the
synthetic control constructed via the procedure outlined above, Panel
C makes this post-tax distinction even clearer. The synthetic control
matches Uganda closely in the entire pre-tax period, but there is
sharp break in trend at the onset of the tax that persists throughout
the post-treatment period.

Panel D of Figure \ref{fig:main_effect} plots the estimated treatment
effect of the social media tax on Twitter use. The estimates suggest
that the daily willingness to pay for Twitter remains below 200 shillings
during most periods for 10 to 20 percent of users. Furthermore, the
estimated effect is consistently outside the .025 and .975 quantiles
of the scaled placebo distribution.

Tweet and user characteristics in the Twitter dataset reveal heterogeneous
effects of the tax. For each category and period, we determine the
number of unique users displaying a characteristic at least once.\footnote{See the Online Appendix for additional details on how these groups
are constructed. These treatment effects incorporate the combined
effect on Twitter use and user characteristics conditional on use
because the user characteristics are not always defined prior to the
tax. However, the user characteristics considered here are unlikely
to be effected by the tax in the timespan examined.} We then re-estimate the synthetic control treatment effects and report
the average of the post-intervention treatment effects along with
the corresponding placebo distributions.

\begin{figure}[H]
\caption{Effect of the Tax on User Composition\label{fig:heterogeneity}}

\begin{centering}
\includegraphics{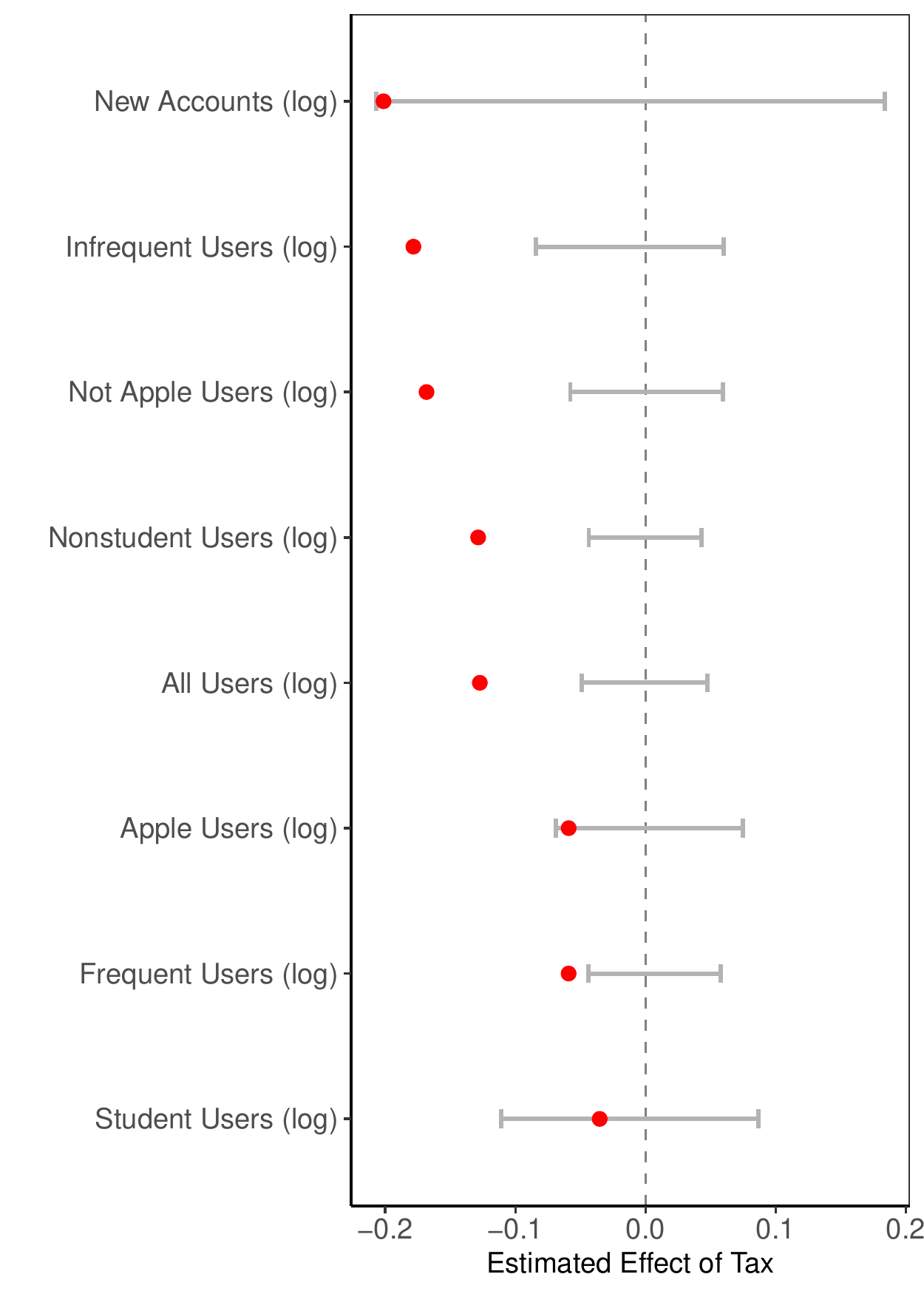}
\par\end{centering}
{\footnotesize{}Notes: The figure plots the average estimated treatment
effect }$\frac{1}{T_{1}+1}\sum_{t\geq0}\hat{\tau}_{0t}${\footnotesize{}
(dot) of the social media tax on various outcomes along with the .025
and .975 quantiles of the scaled placebo distribution $\mathcal{P}$
(grey bars). See the text and Online Appendix for definitions of each
outcome.}{\footnotesize\par}
\end{figure}

Figure \ref{fig:heterogeneity} shows the tax had a larger effect
on newer, less frequent, and poorer users. There was a 20.1 percent
reduction in new account creation as a result of the tax. For users
that tweeted less than one tweet per day on average, there was a 17.8
percent drop. For individuals without an Apple device (an indicator
of wealth), we see a 16.8 percent reduction. In contrast, there was
an estimated 12.7 percent decrease across all users.

To address concerns regarding anticipatory effects and overfitting,
the Online Appendix (Figure \ref{fig:synthetic_control_effects_early_check})
examines the robustness of the main treatment effects to only using
data from more than 100 days before the onset of the tax to fit the
synthetic control. The Online Appendix (Figures \ref{fig:heterogeneity_plac}
and \ref{fig:aggregation}) also shows the results of a falsification
test using pre-tax data only and the estimated treatment effects when
using 1, 7, and 28-day aggregation periods. Lastly, the Online Appendix
(Figure \ref{fig:time_path}) reports the full time path of all treatment
effects considered here and in the section below.

\subsection{Effect of the Tax on Collective Action}

\begin{figure}[H]
\caption{Effect of the Tax on Collective Action\label{fig:collective_action}}

\medskip{}

\begin{centering}
\includegraphics{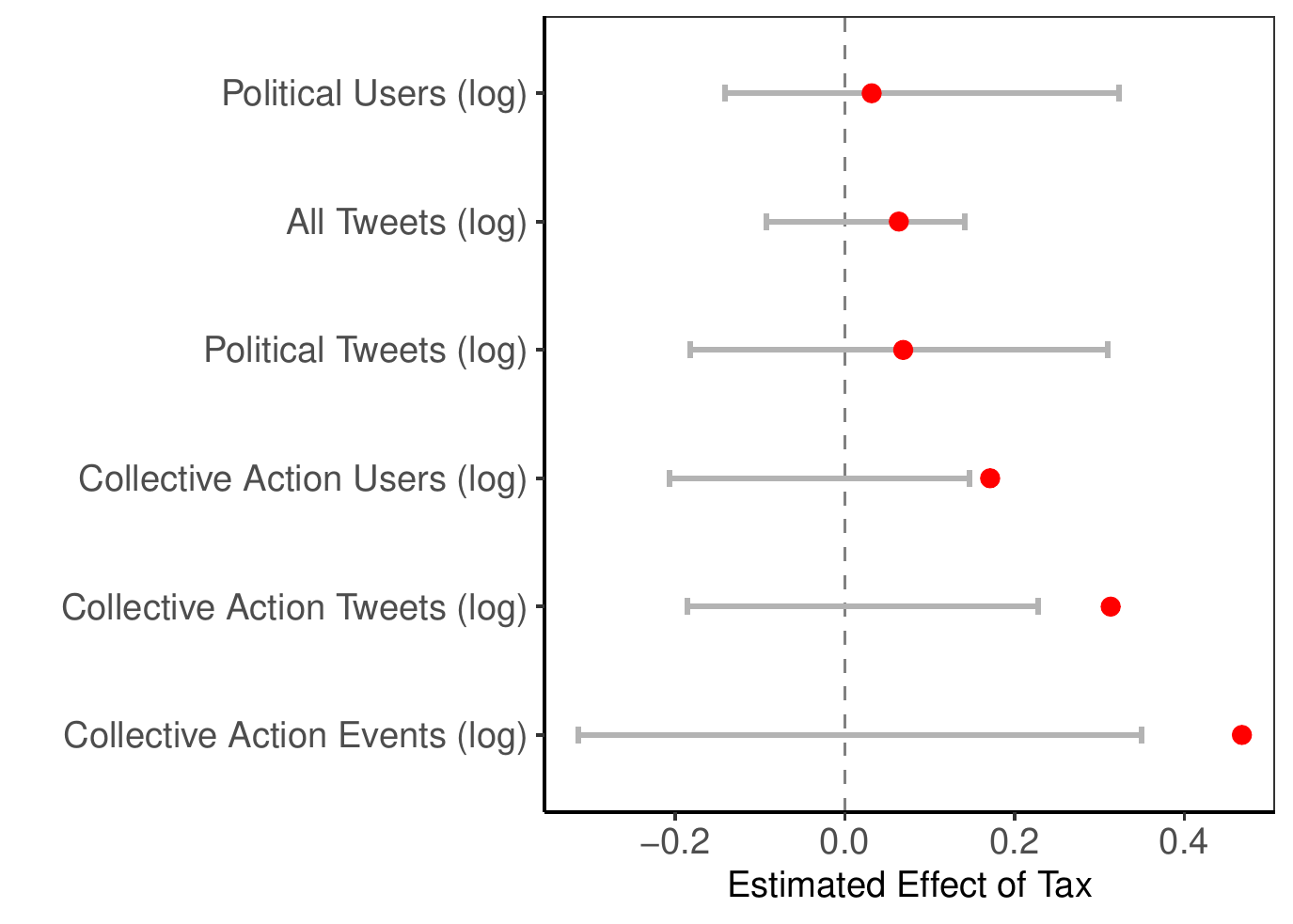}
\par\end{centering}
{\footnotesize{}Notes: The figure plots the average estimated treatment
effect }$\frac{1}{T_{1}+1}\sum_{t\geq0}\hat{\tau}_{0t}${\footnotesize{}
(dot) of the social media tax on various outcomes along with the .025
and .975 quantiles of the scaled placebo distribution $\mathcal{P}$
(grey bars). See the Online Appendix for further details on each outcome.}{\footnotesize\par}
\end{figure}

To examine the impact of the tax on collective action, we first construct
an indicator for whether a tweet contains phrases associated with
collective action (e.g., ``protest'' and ``rally'').\footnote{See the Online Appendix for the complete list of collective action
and political phrases. English is Uganda's only official language,
and there is limited Twitter support for local languages.} Figure \ref{fig:collective_action} shows an estimated 31.3 percent
increase in the number of tweets referencing collective action events
in response to the tax. In contrast, there is no meaningful increase
in the total number of tweets or the number of tweets containing political
phrases unrelated to collective action (e.g., ``president'' and
``government'').

The increase in the amount of online discussion of collective action
is partly driven by an increase in the number of unique users participating
in these discussions. Figure \ref{fig:collective_action} shows a
17.1 percent increase in the number of users that discuss collective
action in a given period. This increase is statistically significant
relative to the placebo distribution.

To examine whether this increase in online discussion of collective
action is associated with an increase in offline activity, we use
data on protests and riots from ACLED and ICEWS. We use the averaged
value for the number of events recorded across both datasets for a
given period. Figure \ref{fig:collective_action} suggests that collective
action events increased by 46.7 percent after the implementation of
the tax\textemdash well outside the .025 and .975 quantiles of the
placebo distribution. The Online Appendix (Figure \ref{fig:time_path},
last panel) shows that the estimated treatment effects on collective
action events are highest in the first 100 days after the tax's implementation.

The Online Appendix (Figure \ref{fig:collective_action_plac}) reports
the results of a falsification test where the synthetic control is
estimated with data more than 100 days prior to the tax's implementation
and the estimated treatment effects are averaged over the 100 day
pre-implementation period. Consistent with the tax driving the effect
on collective action outcomes, the average estimated effects in the
falsification test are either negative or not statistically meaningful.
The Online Appendix (Figure \ref{fig:collective_action_robust}) also
reports sensitivity analysis to alternative specifications of the
outcome variables.

\section{Discussion and Conclusion}

Censorship strategies face a trade off between decreasing access to
sensitive information and creating visible evidence of government
interference in the information environment that has a tendency to
create backlash from those experiencing it.\footnote{Streisand effects can also increase consumption of specific pieces
of sensitive information by drawing attention to and updating beliefs
on what the government does not want citizens to be aware of (Nabi
2014; Jansen and Martin 2015; Roberts 2018).} Backlash is a key reason the Chinese government has shifted towards
more subtle approaches to censorship, including flooding social media
with benign information that distracts from sensitive topics (King
et al. 2017; Roberts 2018). Highly salient forms of digital repression
can also reduce coordination costs for collective action by creating
a focal point or providing common information (Chwe 2000; Tucker 2007;
Egorov et al. 2009; Chwe 2013).

News coverage of Uganda corroborates this mechanism:
\begin{quote}
\begin{singlespace}
{[}the social media tax{]} backfired, and the backlash was fierce.
Young Ugandans, who organised online under \#NoSocialMediaTax and
\#ThisTaxMustGo hashtags, poured into the streets of the capital city
Kampala to express their anger against the government. (Durmaz 2019)
\end{singlespace}
\end{quote}
The language used by opposition leaders on Twitter also suggests the
tax acted as a focal point. Under the handle @HEBobiWine, Kyagulanyi
tweeted the following on July 11, 2018,
\begin{quote}
\begin{singlespace}
..there's no amount of bullets, teargas or arrests that will stop
us. Some of our colleagues have been beaten up and others arrested.
They must be freed for they have nothing against the police but rather
the terrible tax. {[}...{]} \#ThisTaxMustGo.
\end{singlespace}
\end{quote}
The Online Appendix (Figure \ref{fig:tax_and_collective}) shows that
over 30 percent of collective action tweets in the ten days after
the tax's implementation include the string ``tax,'' whereas this
proportion is typically below 5 percent.

In addition to the signalling value of the tax, the effect of the
tax on the composition or structure of the network can alter collective
action outcomes. In network models of collective action, an individuals'
decision to participate often depends on the participation rate of
her local network (Chwe 2000; Watts 2002; Barbera and Jackson 2019).
If tendency for collective action and willingness to pay the tax are
positively correlated, then the tax could increase the likelihood
of successful coordination of collective action by increasing the
density of activists. In the Online Appendix, we develop a model following
Jackson and Yariv (2007), but allowing for endogenous Twitter use,
to highlight the importance of this correlation. Table \ref{tab:who_protests}
in the Online Appendix shows that Apple users and long-time users
are more likely to tweet about collective action in the pre-intervention
period. These users are also users who are more likely to continue
using Twitter after the tax, thus, providing some evidence for the
tax altering the network composition in a way that could facilitate
collective action.

While both mechanisms are consistent with aspects of the data, we
are unable to differentiate which, if any, of the aforementioned mechanisms
are the primary driver for the increase in collective action in the
period after the tax. Furthermore, we can not exclude the possibility
that other concurrent changes are driving the observed trends rather
than the tax itself. However, the anecdotal evidence and language
of elites suggest that the tax was an important impetus to subsequent
collective action.

Overall, the results suggest the economic forces driving selection
into social media use and the signalling value of the tax likely caused
this form of digital repression to backfire by increasing collective
action despite the reduction in social media users.

\end{spacing}

\pagebreak{}

\section*{References}

\leftskip=2em 
\parindent=-2em
\onehalfspacing

\noindent

Abadie, Alberto, Alexis Diamond, and Jens Hainmueller. 2010. Synthetic
control methods for comparative case studies: Estimating the effect
of California's Tobacco Control Program. \emph{Journal of the American
Statistical Association}. 105:493\textendash 505.

Al Jazeera. 2019. Ugandan pop star opposition MP Bobi Wine arrested
again. \emph{Al Jazeera}. Accessed at \href{https://www.aljazeera.com/news/2019/04/ugandan-pop-star-opposition-mp-bobi-wine-arrested-190429153709250.html}{https://www.aljazeera.com/news/2019/04/ugandan-pop-star-opposition-mp-bobi-wine-arrested-190429153709250.html}.

Ananyev, Maxim, Galina Zudenkova, and Maria Petrova. 2017. Information
and communication technologies, protests, and censorship. \emph{Working
Paper}. Accessed at \href{https://ssrn.com/abstract\%3D2978549}{https://ssrn.com/ abstract=2978549}.

Anderson, Simon P. and Andre de Palma. 2009. Information congestion.
\emph{RAND Journal of Economics}. 40(4): 688\textendash 709.

Barbera, Salvador and Matthew O. Jackson. 2019. A model of protests,
revolution, and information. \emph{Working Paper}. Accessed at \href{https://ssrn.com/abstract\%3D2732864}{https://ssrn.com/abstract=2732864}.

BBC. 2017. Uganda MPs vote to scrap presidential age limit. \emph{BBC.
}Accessed at \href{https://www.bbc.com/news/world-africa-42434809}{https://www. bbc.com/news/world-africa-42434809}.

Boschee, Elizabeth, Jennifer Lautenschlager, Sean O'Brien, Steve Shellman,
James Starz, and Michael Ward. 2015. ICEWS Coded Event Data. Harvard
Dataverse Network. \href{http://dx.doi.org/10.7910/DVN/28075}{http://dx. doi.org/10.7910/DVN/28075}.

Bursztyn, Leonardo, Davide Cantoni, David Y. Yang, Noam Yuchtman,
and Y. Jane Zhang. 2019. Persistent political engagement: Social interactions
and the dynamics of protest movements. \emph{Working Paper}. Accessed
at \href{https://home.uchicago.edu/bursztyn/Persistent_Political_Engagement_July2019.pdf}{https://home.uchicago.edu/bursztyn/ Persistent\_Political\_Engagement\_July2019.pdf}.

Chen, Yuyu, and David Y. Yang. Forthcoming. The impact of media censorship:
1984 or brave new world? \emph{American Economic Review}.

Chwe, Michael Suk-Young. 2000. Communication and coordination in social
networks. \emph{Review of Economic Studies}. 67: 1\textendash 16.

Chwe, Michael Suk-Young. 2013. \emph{Rational Ritual: Culture, Coordination,
and Common Knowledge}. Princeton University Press: Princeton, NJ.

Crabtree, Justina. 2018. Facebook says it is committed to Uganda despite
social media tax to quash `gossip'. \emph{cnbc.com}. Accessed at \href{https://www.cnbc.com/2018/08/10/facebook-committed-to-uganda-despite-social-media-tax.html}{https://www.cnbc.com/2018/08/10/facebook-committed-to-uganda-despite-social-media-tax.html}
on April 3, 2019.

Durmaz, Mucahid. 2019. Do African autocrats fear the internet? \emph{TRT
World}. Accessed at \href{https://www.trtworld.com/magazine/do-african-autocrats-fear-the-internet-23327}{https://www.trtworld.com/magazine/do-african-autocrats-fear-the-internet-23327}.

Egorov, Georgy, Sergei Guriev, and Konstantin Sonin. 2009. Why resource-poor
dictators allow freer media: A theory and evidence from panel data.
\emph{American Political Science Review}. 103(4): 645\textendash 668.

Enikolopov, Ruben, Alexey Makarin, Maria Petrova, and Leonid Polishchuk.
2017. Social image, networks, and protest participation. \emph{Working
Paper}\@. Accessed at \href{https://ssrn.com/abstract\%3D2940171}{https://ssrn.com/ abstract=2940171}.

Enikolopov, Ruben, Alexey Makarin, and Maria Petrova. 2018. Social
media and protest participation: Evidence from Russia. \emph{Working
Paper}. Accessed at \href{https://ssrn.com/abstract\%3D2696236}{https://ssrn.com/abstract= 2696236}.

Gohdes, Anita. 2015. Pulling the plug: Network disruptions and violence
in civil conflict. \emph{Journal of Peace Research}. 52(3): 352\textendash 367.

Gould, Roger V. 1993. Collective action and network structure. \emph{American
Sociological Review}. 58(2): 182\textendash 196.

Griffiths, James. 2019. Democratic Republic of Congo internet shutdown
shows how Chinese censorship tactics are spreading. \emph{CNN. }Accessed
at \href{https://www.cnn.com/2019/01/02/africa/congo-internet-shutdown-china-intl/index.html}{https://www.cnn.com/2019/01/02/ africa/congo-internet-shutdown-china-intl/index.html}.

Hassanpour, Navid. 2014. Media disruption and revolutionary unrest:
Evidence from Mubarak's quasi-experiment. \emph{Political Communication}.
31(1):1\textendash 24.

Hobbs, William and Margaret E. Roberts. 2018. How sudden censorship
can increase access to information. \emph{American Political Science
Review}. 112(3):1\textendash 16.

Iyer, Ganesh and Zsolt Katona. 2016. Competing for attention in social
communication markets. \emph{Management Science}. 62(8): 2304\textendash 2320.

Jackson, Matthew O. and Leeat Yariv. 2007. Diffusion of Behavior and
Equilibrium Properties in Network Games. \emph{American Economic Review}
\emph{(Papers and Proceedings)}. 97(2): 92\textendash 98

Jansen, Sue Curry and Brian Martin. 2015. The Streisand effect and
censorship backfire. \emph{International Journal of Communication}.
9: 656\textendash 671.

Kagumire, Rosebell. 2018. Bobi Wine and the beginning of the end of
Museveni's power. \emph{Al Jazeera. }Accessed at \href{https://www.aljazeera.com/indepth/opinion/bobi-wine-beginning-museveni-power-180828111608108.html}{https://www.aljazeera.com/indepth/opinion/bobi-wine-beginning-museveni-power-180828111608108.html}.

King, Gary, Jennifer Pan, and Margaret E. Roberts. 2013. How censorship
in China allows government criticism but silences collective expression.
\emph{American Political Science Review}. 107(2): 1\textendash 18.

\textemdash . 2014. Reverse-engineering censorship in China: randomized
experimentation and participant observation. \emph{Science}. 345(6199):
1\textendash 10.

\textemdash . 2017. How the Chinese government fabricates social media
posts for strategic distraction, not engaged argument. \emph{American
Political Science Review}. 111(3): 484\textendash 501.

Leetaru, Kalev H., Shaowen Wang, Guofeng Cao, Anand Padmanabhan, and
Eric Shook. 2013. Mapping the global Twitter heartbeat: The geography
of Twitter. \emph{First Monday}. 18(5-6).

Little, Andrew T. 2016. Communication technology and protest. \emph{Journal
of Politics}. 78(1): 152\textendash 166.

Manacorda, Marco and Andrea Tesei. 2016. Liberation technology: Mobile
phones and political mobilization in Africa. \emph{Working Paper.
}Accessed at \href{https://ssrn.com/abstract\%3D2795957}{https://ssrn.com/abstract=2795957}.

Mugerwa, Yasiin. 2017. China to help Uganda fight internet abuse.
\emph{Daily Monitor}. Accessed at \href{https://www.monitor.co.ug/News/National/China-Uganda-Internet-Evelyn-Anite-Africa-Internet-Users/688334-4032626-u1l61r/index.html}{https://www.monitor.co.ug/News/National/China-Uganda-Internet-Evelyn-Anite-Africa-Internet-Users/688334-4032626-u1l61r/index.html}.

Nabi, Zubair. 2014. Censorship is futile. \emph{First Monday}. 19(11).

Observer. 2017. Uganda to launch own version of Facebook, Twitter
- UCC. \emph{The Observer}. Accessed at \href{https://observer.ug/news/headlines/56506-uganda-to-launch-own-version-of-facebook-next-year-ucc.html}{https://observer.ug/news/headlines/56506-uganda-to-launch-own-version-of-facebook-next-year-ucc.html}.

Oliver, Pamela E. and Gerald Marwell. 1988. The paradox of group size
in collective action: A theory of the critical mass. II. \emph{American
Sociological Review}. 53(1): 1\textendash 8.

Pan, Jennifer and Alexandra A. Siegel. 2018. Physical repression and
online dissent: Evidence from Saudia Arabia. \emph{Working Paper.
}Accessed at \href{https://alexandra-siegel.com/wp-content/uploads/2018/11/Pan_Siegel_Saudi_Twitter_Oct29.pdf}{https://alexandra-siegel.com/wp-content/uploads/2018/11/Pan\_Siegel\_Saudi\_Twitter\_Oct29.pdf}.

Qin, Bei, David Stromberg, and Yanhui Wu. 2017. Why does China allow
freer social media? Protests versus surveillance and propaganda. \emph{Journal
of Economic Perspectives}. 31(1): 117\textendash 140.

Raleigh, Clionadh, Andrew Linke, Håvard Hegre and Joakim Karlsen.
2010. \textquotedblleft Introducing ACLED\textendash Armed conflict
location and event data.\textquotedblright{} \emph{Journal of Peace
Research} 47(5): 65\textendash 660.

Roberts, Margaret E. 2018. \emph{Censored: Distraction and Diversion
Inside China's Great Firewall}. Princeton University Press: Princeton,
NJ.

Siegel, David A. 2009. Social networks and collective action. \emph{American
Journal of Political Science}. 53(1): 122\textendash 138.

Steinert-Threlkeld, Zachary. 2018. \emph{Twitter as Data}. Cambridge
University Press: Cambridge, UK.

Sunstein, Cass R. Forthcoming. Valuing Facebook. \emph{Behavioural
Public Policy}.

Tucker, Joshua A. 2007. Enough! Electoral fraud, collective action
problems, and post-communist colored revolutions. \emph{Perspectives
on Politics}. 5(3): 535\textendash 551.

Van Zandt, Timothy. 2004. Information overload in a network of targeted
communication. \emph{RAND Journal of Economics}. 35(3): 542\textendash 560.

Watts, Duncan J. 2002. A simple model of global cascades on random
networks. \emph{Proceedings of the National Academy of Sciences.}
99(9): 5766\textendash 5771.

\parindent=2em
\leftskip=0em

\pagebreak{}

\clearpage

\newpage\renewcommand\thesection{A\arabic{section}} 
\renewcommand{\thesubsection}{A\arabic{section}.\arabic{subsection}} 
\setcounter{section}{0}  
\setcounter{table}{0} 
\renewcommand{\thetable}{A\arabic{table}} 
\setcounter{figure}{0} 
\renewcommand{\thefigure}{A\arabic{figure}}

\section{Online Appendix}

\subsection{Additional Data Details\label{subsec:Additional-Data-Details}}

See replication code for exact details on the data construction.

\subsubsection{Twitter Outcomes}

For the Twitter outcomes, we restrict the set of countries for which
we use to construct the synthetic control to those in the top 80 percent
of the sample of sub-Saharan African countries with respect to the
average number of unique, active users per period, and we do this
separately for each level of temporal aggregation. We also exclude
tweets whose user description contains phrases associated with bot
activity.\footnote{``weather,'' ``4:20,'' ``job,'' ``career,'' ``hire,'' and
``hiring.''} For all logged variables, we use the log of one plus the level.
\begin{itemize}
\item New accounts: The number of users in our dataset who created their
Twitter account in a given period.
\item Infrequent users: Users who, upon their first appearance in our dataset,
averaged less than one tweet per day as measured by the number of
posted statuses listed on their account and the date of account creation.
\item Not Apple users: Users for which our dataset never contains a tweet
posted via ``ios,'' ``ipad,'' or ``iphone'' in a given period.
\item Student users: Users for which the user description or user location
includes ``university,'' ``college,'' ``universite,'' ``universita,''
``universidade,'' ``faculdade'' or ``student'' for at least
one tweet in our dataset in a given period.
\item Activist users: Users for which at least one tweet contains a phrase
referencing collecting action in a given period.\footnote{`protest', `rally', `rallies', ` freedom of assembly ', ` freedom
of expression ', ` riot', `march against', `march with', `inciting
violence', `unlawful assembly', `civil unrest', `demonstration', `i
stand with', `take to the street', `taking to the street', or `boycott'}
\item Political users: Users for which at least one tweet contains a phrase
referencing the government in a given period.\footnote{`supreme court', `legislator', `election', `president', `vote', `ballot',
`administracao', `presidente', ` mp ', `constitution', `partisan',
` mps ', `administration', `dictator', `gouvernement', `citizen',
`republic', `political party', `political parties', `parlement', `parliament',
`government', ` law ', `constituent', `governo', `democrac', `politic',
`amministrazione', `military', `constituency', and `parlamento'.}
\item Collective action tweets: Tweets containing at least one phrase referencing
collective action as defined above for activist users.
\end{itemize}

\subsubsection{ACLED and ICEWS Outcomes}

The ACLED data includes all countries in Africa and is described by
Raleigh et al. (2010). It was downloaded from \href{https://www.acleddata.com/}{https://www.acleddata.com/}
on March 6, 2019. We restrict attention to events labelled as ``Riots/protests.''
Other event categories include, among others, ``Violence against
citizens'' and ``Battle-No change of territory.''

The ICEWS data comes from Boschee et al. (2015). It is a machine-coded
dataset that generates 20 event types, one of which is protests. Of
the 45 events in Uganda during the main window of analysis, none are
exact duplicates.\footnote{Defined by source, actor, target, and location.}

Separately, for each dataset, we compute the number of events that
occur in a given time period. We then take the average across both
datasets as our baseline measure of the number of collective action
events for a given country in a given period, where this measure is
appropriately transformed depending on the specification (e.g., log).
We restrict attention to countries that appear in both our ACLED and
ICEWS samples throughout. For all logged variables, we use the log
of one plus the level.

\clearpage

\subsection{Additional Empirical Details}

To estimate the impact of the social media tax, we adopt a synthetic
control approach (Abadie et al. 2010). Let $Y_{it}(1)$, $Y_{it}(0)$
be outcome $Y$ for country $i$ at time $t$ with and without the
social media tax respectively. Let $i=0$ represent Uganda, $i\in\{1,...,N\}$
represent the control countries, and $t\in\{-T_{0},...,0,...,T_{1}\}$
be the timespan considered where $0$ indicates the first period in
which the social media tax was implemented (July 1, 2018). Each time
period represents a 10-day span. 

Assume 
\begin{equation}
Y_{it}(0)=\delta_{t}+\lambda_{t}\cdot\mu_{i}+\epsilon_{it}\label{eq:factor-1}
\end{equation}
where $\delta_{t}$ is a constant factor across countries, $\mu_{i}$
is a vector of unobserved factors for each country $i$ with time-varying
loadings $\lambda_{t}$, and $\epsilon_{it}$ is a mean zero error
term. The treatment effect of the tax $\tau$ is assumed to be additive,
but can vary across countries and time 
\begin{equation}
Y_{it}(1)=Y_{it}(0)+\tau_{it}.
\end{equation}
Under mild conditions,\footnote{See Abadie et al. (2010).} if a vector
of weights $w$, where $\sum_{j=1}^{N}w_{j}=1$ and $w_{j}\geq0$
for all $j\in\{1,...,N\}$, is chosen such that $\sum_{j=1}^{N}w_{j}Y_{jt}(0)=Y_{0t}(0)$
holds approximately for all $t<0$ when $T_{0}$ is relatively large,
then 
\begin{equation}
\hat{\tau}_{0t}=Y_{0t}(1)-\sum_{j=1}^{N}\hat{w}_{j}Y_{jt}(0)\label{eq:estimator-1}
\end{equation}
provides an unbiased estimate of $\tau_{0t}$.\footnote{In estimating the weights, we choose $w$ to minimize $||{\bf X_{0}}-{\bf X_{1}}w||=\sqrt{({\bf X_{1}}-{\bf X_{0}}w)'{\bf I}({\bf X_{1}}-{\bf X_{0}}w)}$
where ${\bf X_{0}}=(Y_{0,T_{0}}(0),...,Y_{0,-1}(0))'$, ${\bf X}_{1}$
is an analogous matrix over the control units, and ${\bf I}$ is an
identity matrix. We estimate the weights separately for each outcome
of interest.}

For inference, we compare the estimated treatment effect for Uganda
$\hat{\tau}_{0t}$ to the placebo distribution of estimated effects
created by repeating the synthetic control procedure for each control
unit. Specifically, we estimate $\tau_{it}$ for each control unit
and compute the associated mean squared prediction error 
\[
\sigma_{i}^{2}=\frac{1}{T_{0}}\sum_{t<0}(Y_{it}(0)-\hat{Y}_{it}(0))^{2}.
\]
We then compare $\hat{\tau}_{0t}$ to the scaled placebo distribution
$\mathcal{P}_{t}=\{\hat{\tau}_{1}\frac{\sigma_{0}}{\sigma_{1}},...,\hat{\tau}_{N}\frac{\sigma_{0}}{\sigma_{N}}\}$.
Scaling by $\frac{\sigma_{0}}{\sigma_{i}}$ accounts for systematic
differences in how well the synthetic control performs across countries
(e.g., a country on the edge of the convex hull will tend to perform
poorly). An estimate of $\hat{\tau}_{0t}$ outside the .025 and .975
quantiles of $\mathcal{P}_{t}$ is indicative of a statistically meaningful
effect. Note that these are point-wise distributions.

See replication code for exact details on the estimation procedures.

\subsection{Model of Collective Action with Social Media Tax}

The following discussion builds on Jackson and Yariv (2007). There
is a continuum of agents. Each agent decides whether to pay a price
$q$ and join Twitter. Users have an intrinsic valuation of Twitter
$w_{i}\geq0$ with c.d.f. $F_{w}$.

Each agent subsequently chooses whether to participate in a protest
and has costs of participating $c_{i}\geq0$. The utility from participating
in the protest is a function of the average participation of connections
on Twitter $v(x)$ (where Twitter connections are formed randomly
from the population of Twitter users), and that $v(\cdot)$ is an
increasing function in $x$.\footnote{Define $x\equiv0$ for non-Twitter users and $v(0)=0.$}

Let $c_{i}$ and $w_{i}$ be jointly normal with mean $\mu$ and variance
$\Omega$. Agents will choose to protest if $v(x)>c_{i}$. Let $F_{c|w_{i}\geq q-v(x)}$
denote the conditional distribution of idiosyncratic protest costs
for users on Twitter for a given rate of protest participation $x$.
Following Jackson and Yariv (2007), we have that 
\[
x=\phi(x)\equiv F_{c|w_{i}\geq q-v(x)}[v(x)]
\]
characterizes the symmetric Bayesian equilibria of the game.

This leads to the following theorem.
\begin{thm}
\textbf{Impact of tax depends on the covariance $\Omega_{cw}$ between
Twitter valuations $w_{i}$ and protest costs $c_{i}$.}
\begin{enumerate}
\item If $q'>q$ and $\Omega_{cw}<0$, then $q'$ will generate more diffusion
of behavior on the Twitter network.
\item If $q'>q$ and $\Omega_{cw}>0$, then $q'$ will generate less diffusion
of behavior on the Twitter network.
\end{enumerate}
\begin{proof}
Note that $F_{c|w_{i}\geq q-v(x)}[v(x)]$ is an increasing {[}decreasing{]}
function in $q$ for all $x$ when $\Omega_{cw}<0$ {[}$\Omega_{cw}>0${]}.
Therefore, $\phi(x,q')>\phi(x,q)$ {[}$\phi(x,q')<\phi(x,q)${]} for
all $x$.

Proposition 1 of Jackson and Yariv (2007) states that, for two regular
environments,\footnote{See Jackson and Yariv (2007) for technical details.}
an environment $\tilde{\phi}(\cdot)$ will generate greater diffusion
of behavior\footnote{In the sense that tipping points are lower and stable equilibria are
higher.} than $\phi(\cdot)$ if $\tilde{\phi}(x)\geq\phi(x)$ for all $x$.
\end{proof}
\end{thm}

\subsection{Online Appendix Figures and Tables}

\begin{figure}[H]
\caption{Robustness to Restricting Model Fitting Period \label{fig:synthetic_control_effects_early_check}}

\medskip{}

\begin{centering}
\textit{\small{}Panel A: Estimated Treatment Effects}{\small\par}
\par\end{centering}
\begin{centering}
\textit{\small{}\includegraphics[scale=0.75]{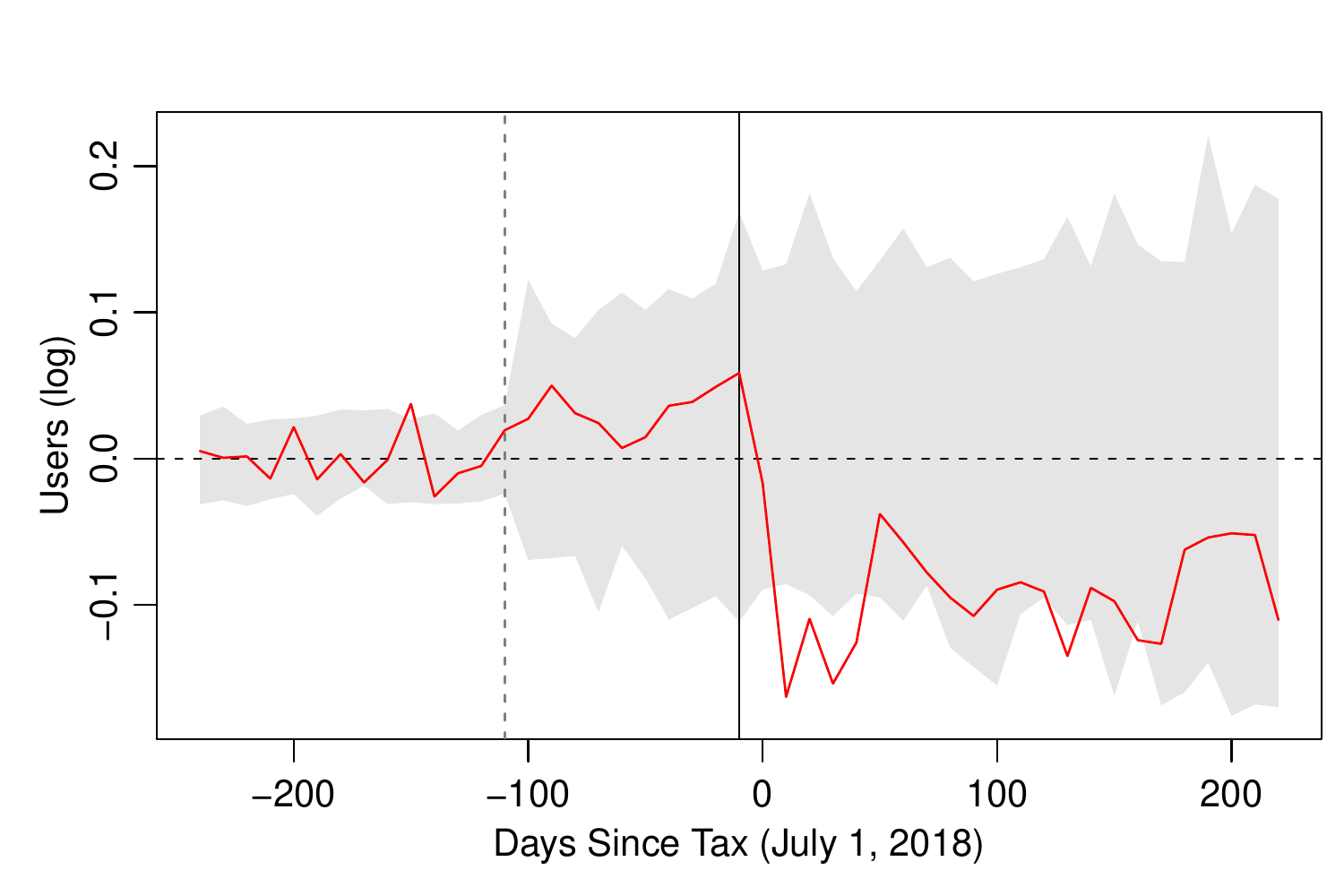}\medskip{}
}{\small\par}
\par\end{centering}
\begin{centering}
\textit{\small{}Panel B: Discussion of Social Media Tax}{\small\par}
\par\end{centering}
\begin{centering}
\includegraphics[scale=0.75]{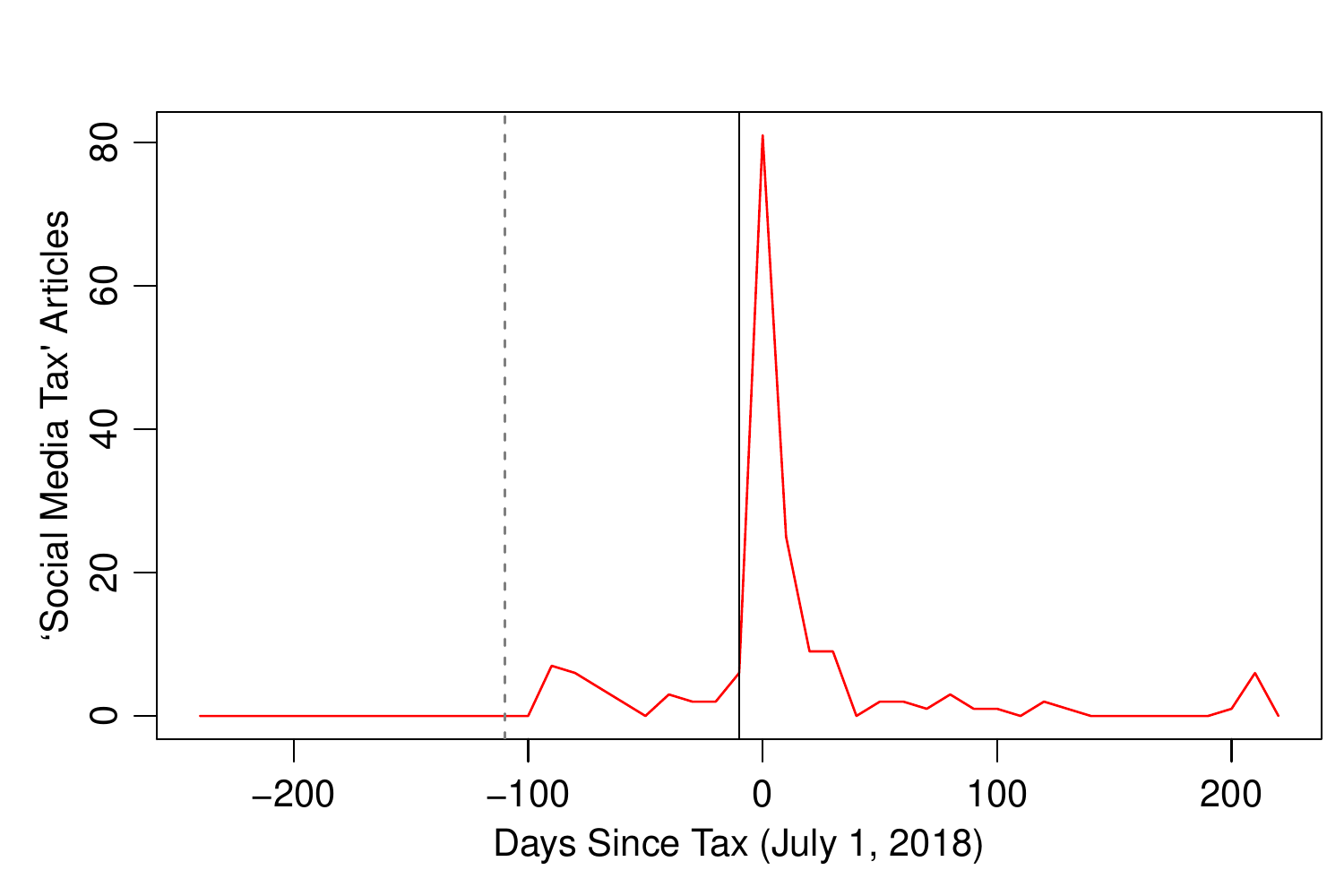}
\par\end{centering}
{\footnotesize{}Notes: Panel A shows the log of one plus the number
of users per period in each 10-day period for Uganda relative to the
synthetic control. The solid vertical line indicates the 10 day period
immediately prior to when the social media tax was implemented in
Uganda (July 1, 2018). The dashed vertical line indicates the last
period which was used to fit the synthetic control. Panel B shows
the total number of news articles, as measured by data from BuzzSumo,
referencing ``social media tax'' with a .ug domain in each 10-day
period.}{\footnotesize\par}
\end{figure}

\begin{figure}[H]
\caption{Effect of the Tax on User Composition \textendash{} Falsification
Test\label{fig:heterogeneity_plac}}

\begin{centering}
\includegraphics{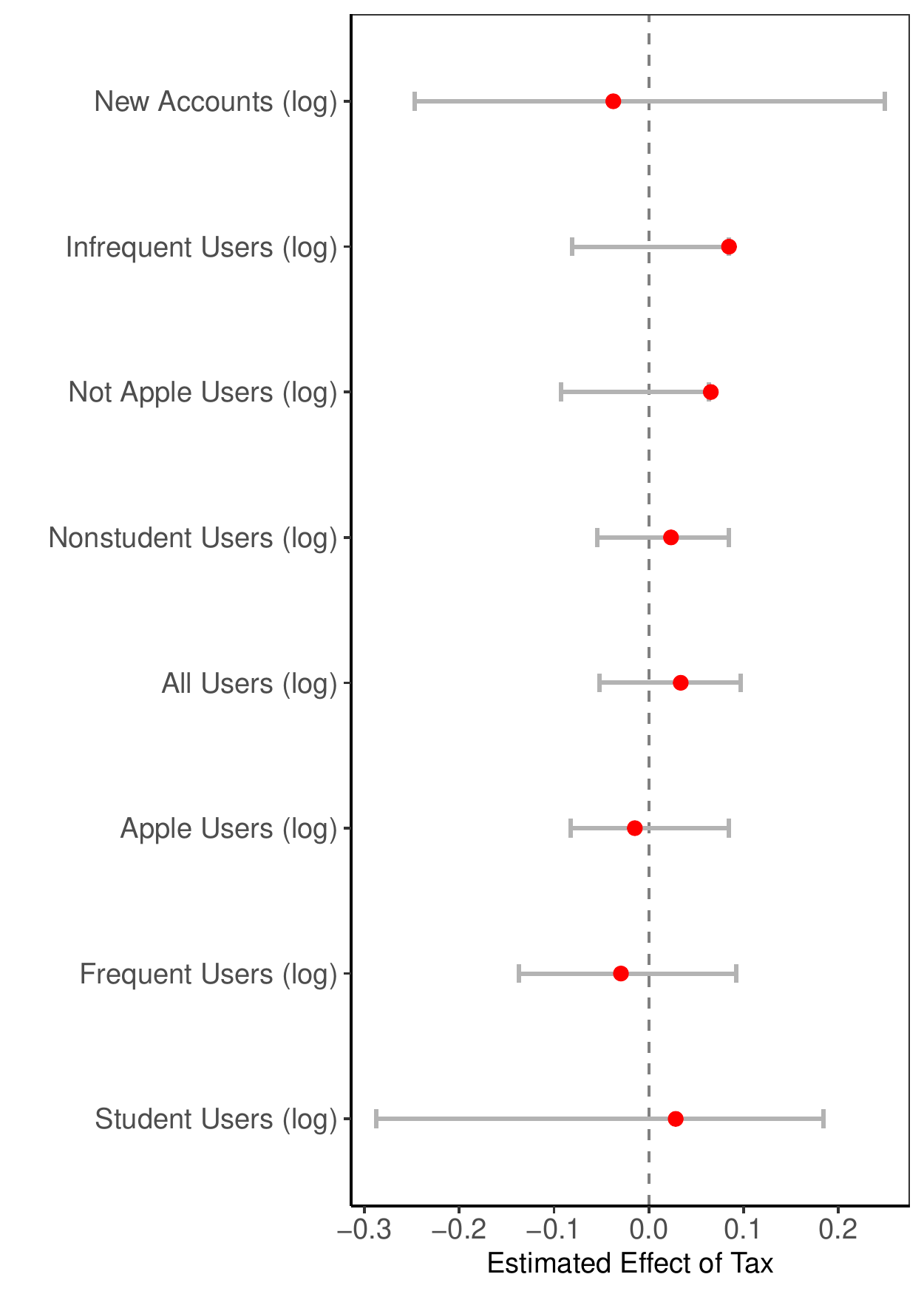}
\par\end{centering}
{\footnotesize{}Notes: The figure plots the average estimated treatment
effect (dot) of the social media tax on various outcomes along with
the .025 and .975 quantiles of the scaled placebo distribution $\mathcal{P}$
(grey bars) when restricting the model fit of the synthetic control
to more than 100 days prior to the tax. The average of the estimated
treatment effects are then taken over the 100 days prior to the tax.
See Section \ref{subsec:Additional-Data-Details} for further details
on each outcome and Figure \ref{fig:heterogeneity} to compare to
the true estimated treatment effects.}{\footnotesize\par}
\end{figure}

\begin{figure}[H]
\caption{Robustness to Other Levels of Aggregation\label{fig:aggregation}}

\medskip{}

\begin{centering}
\emph{Panel A: 1-day Aggregation}
\par\end{centering}
\begin{centering}
\includegraphics[scale=0.5]{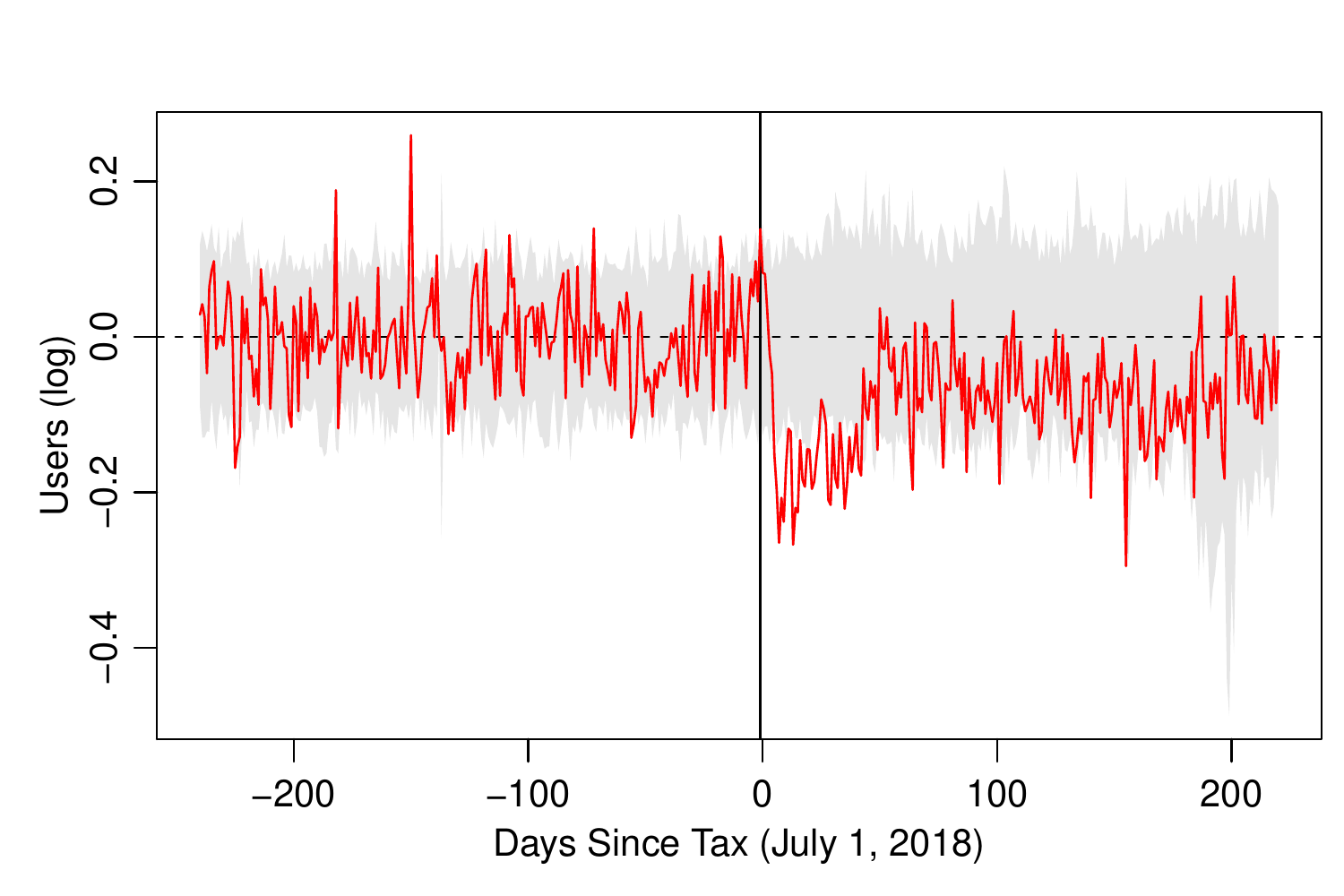}
\par\end{centering}
\medskip{}

\begin{centering}
\emph{Panel B: 7-day Aggregation}
\par\end{centering}
\begin{centering}
\includegraphics[scale=0.5]{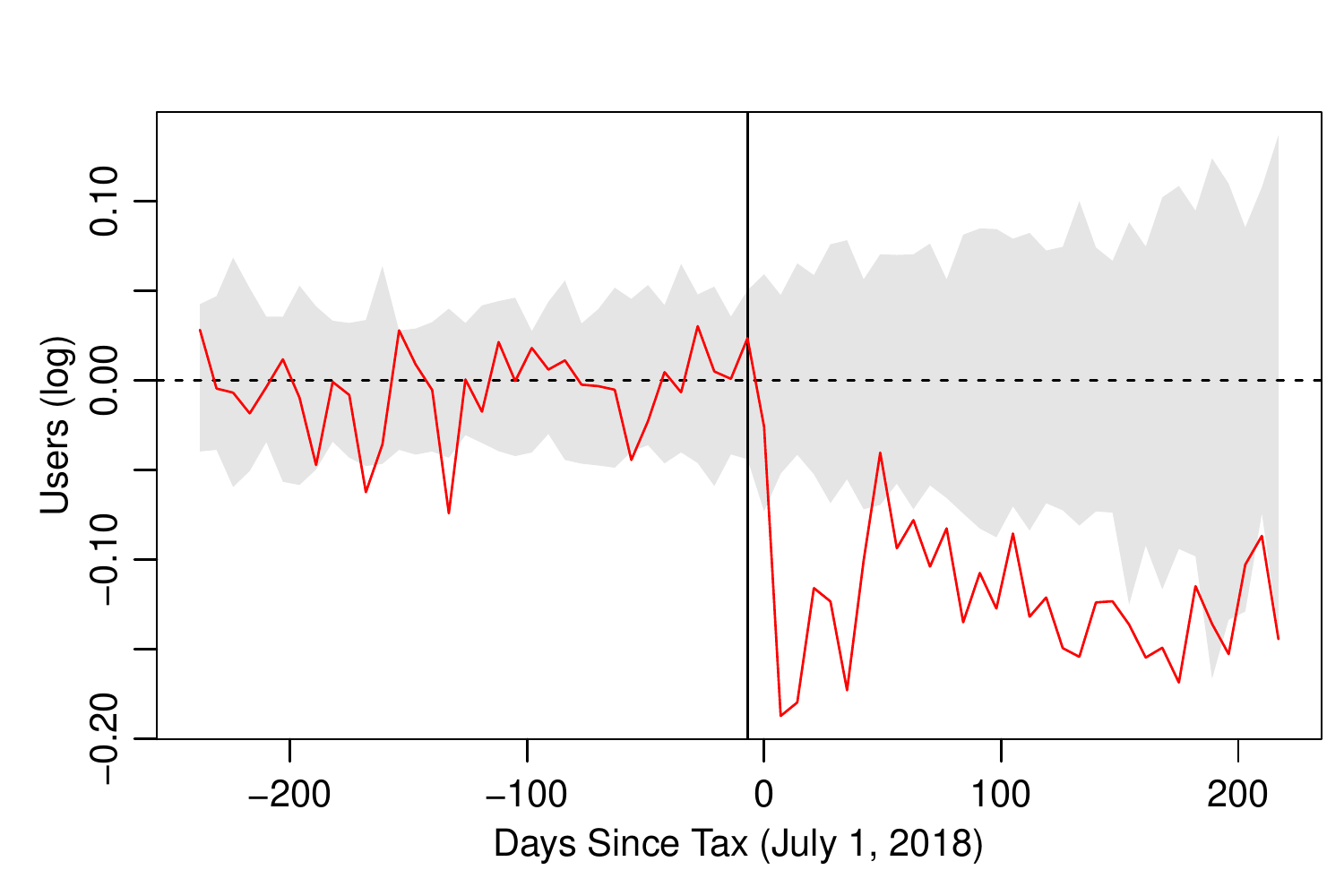}\medskip{}
\par\end{centering}
\begin{centering}
\emph{Panel C: 28-day Aggregation}
\par\end{centering}
\begin{centering}
\includegraphics[scale=0.5]{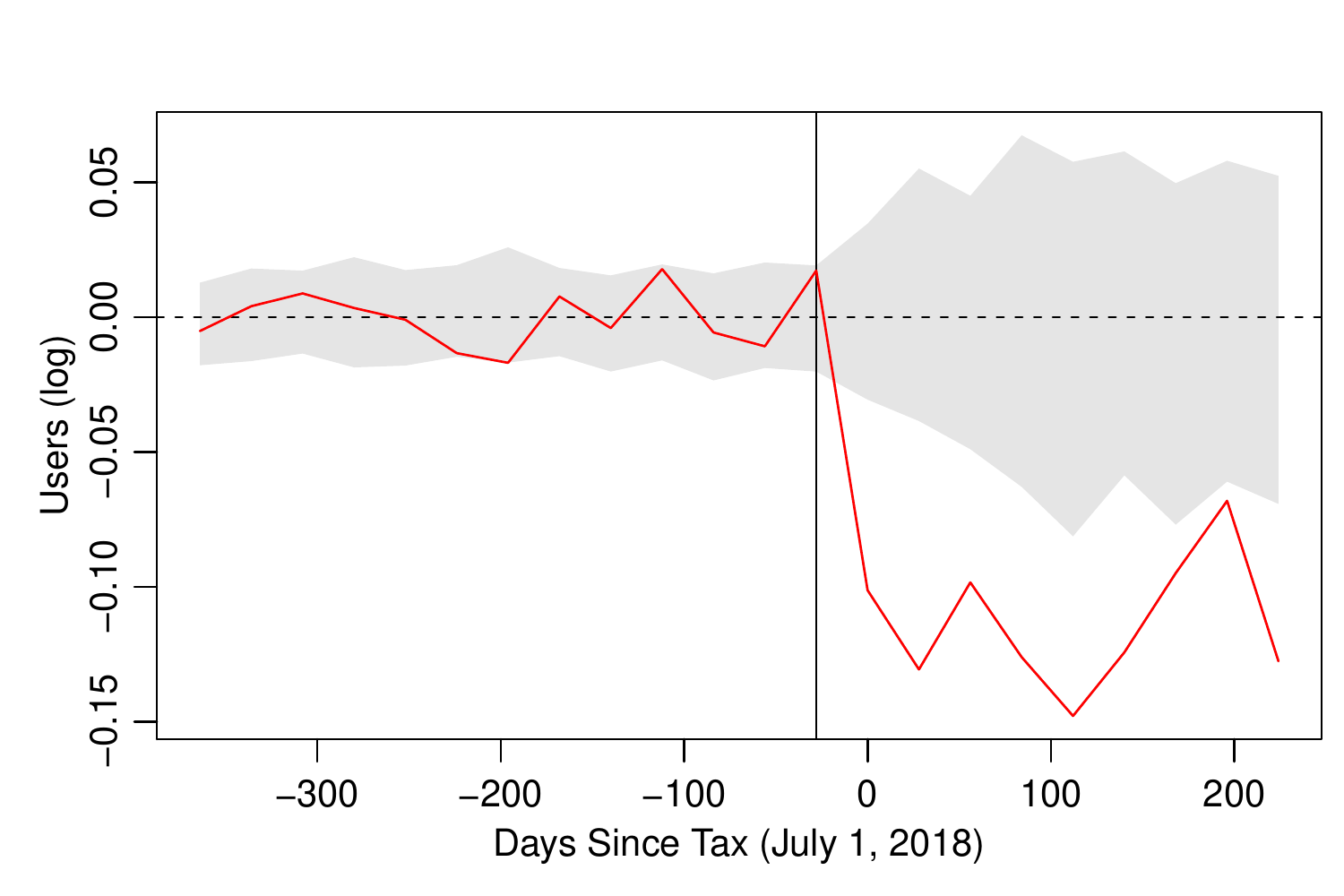}
\par\end{centering}
{\footnotesize{}Notes: The figure shows the estimated effect of the
tax on the log of one plus the number of unique, active users in a
given period using various levels of temporal aggregation. Panel A
uses 1-day periods, Panel B uses 7-day periods, and Panel C uses 28-day
periods. Panel A uses every 10th period in the pre-tax period for
${\bf X_{0}}$ in estimating the weights. Panel B uses every 4th period
in the pre-tax period for ${\bf X_{0}}$ in estimating the weights.
Panel C uses all periods in the pre-tax period for ${\bf X_{0}}$
in estimating the weights. Since Panels A and B do not use all pre-intervention
periods for ${\bf X}_{0}$, the weights are chosen to minimize $\sqrt{({\bf X_{1}}-{\bf X_{0}}w)'{\bf V}({\bf X_{1}}-{\bf X_{0}}w)}$
where ${\bf V}$ is a positive definite diagonal matrix chosen to
minimize the mean squared prediction error across all pre-intervention
periods. The shaded region indicates the .025 and .975 quantiles of
the scaled placebo distribution $\mathcal{P}_{t}$. The vertical line
indicates the 10 day period immediately prior to when the social media
tax was implemented in Uganda (July 1, 2018).}{\footnotesize\par}
\end{figure}

\begin{figure}[H]
\caption{Time Path of Estimated Treatment Effects\label{fig:time_path}}

\includegraphics[scale=0.33]{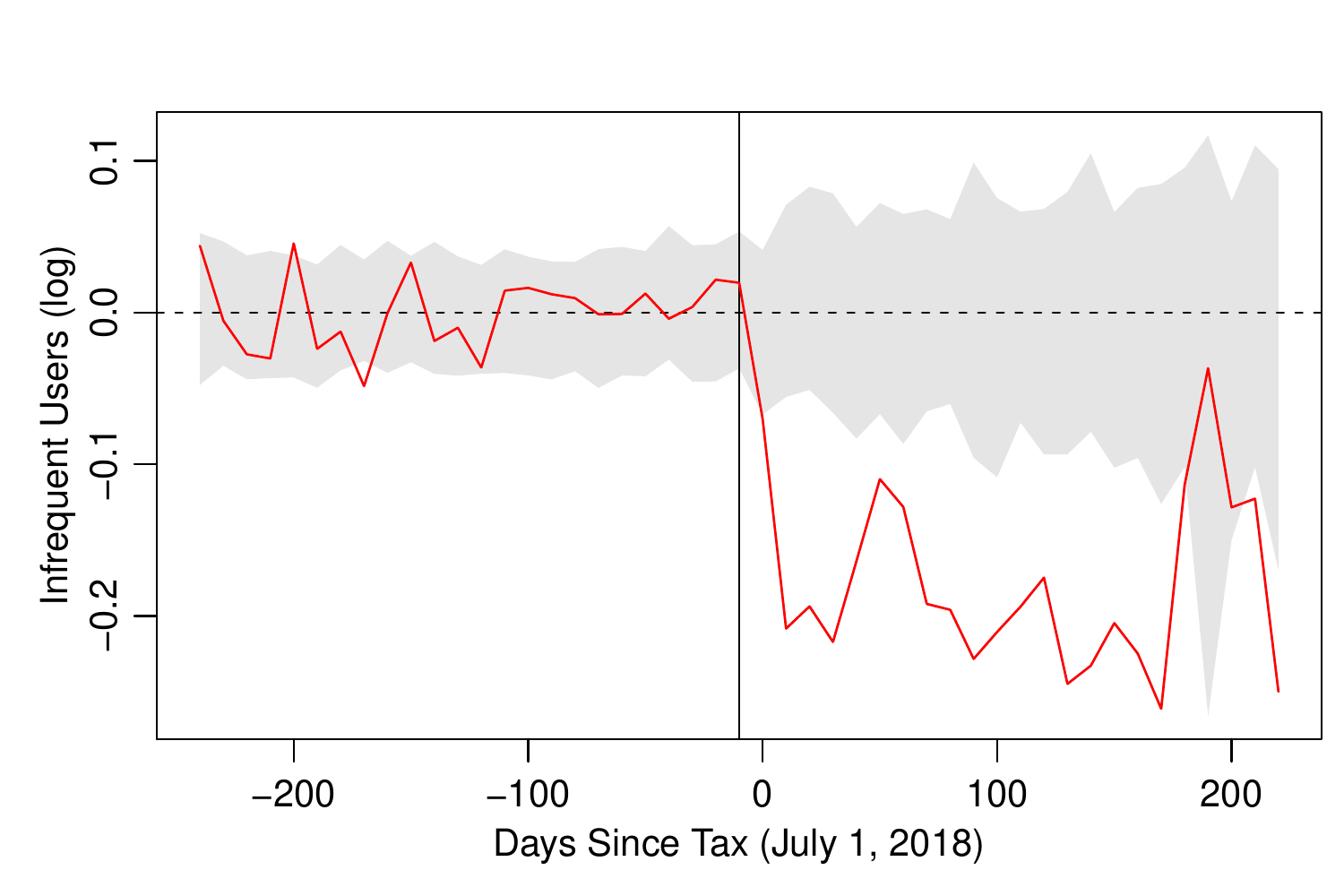}\includegraphics[scale=0.33]{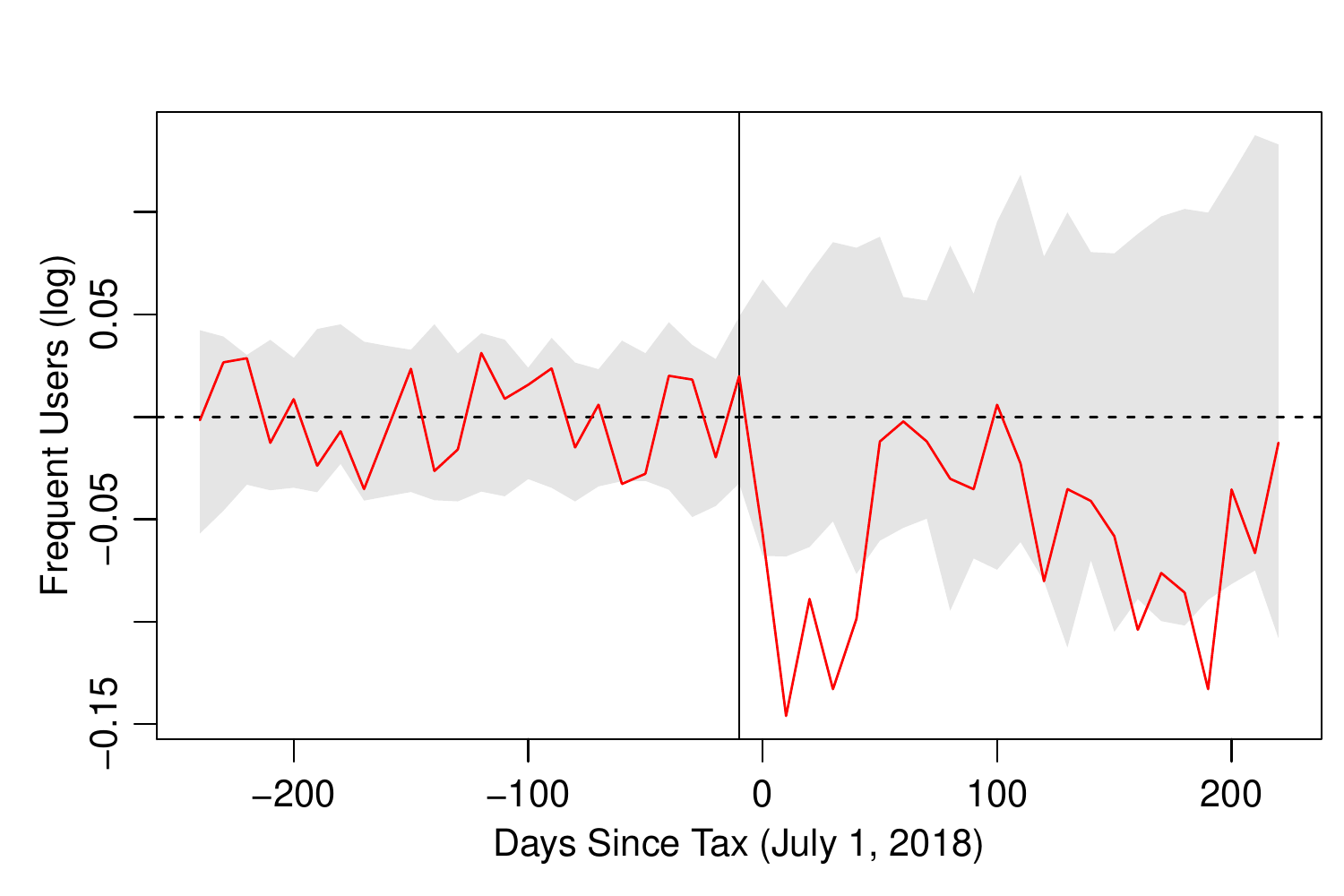}\includegraphics[scale=0.33]{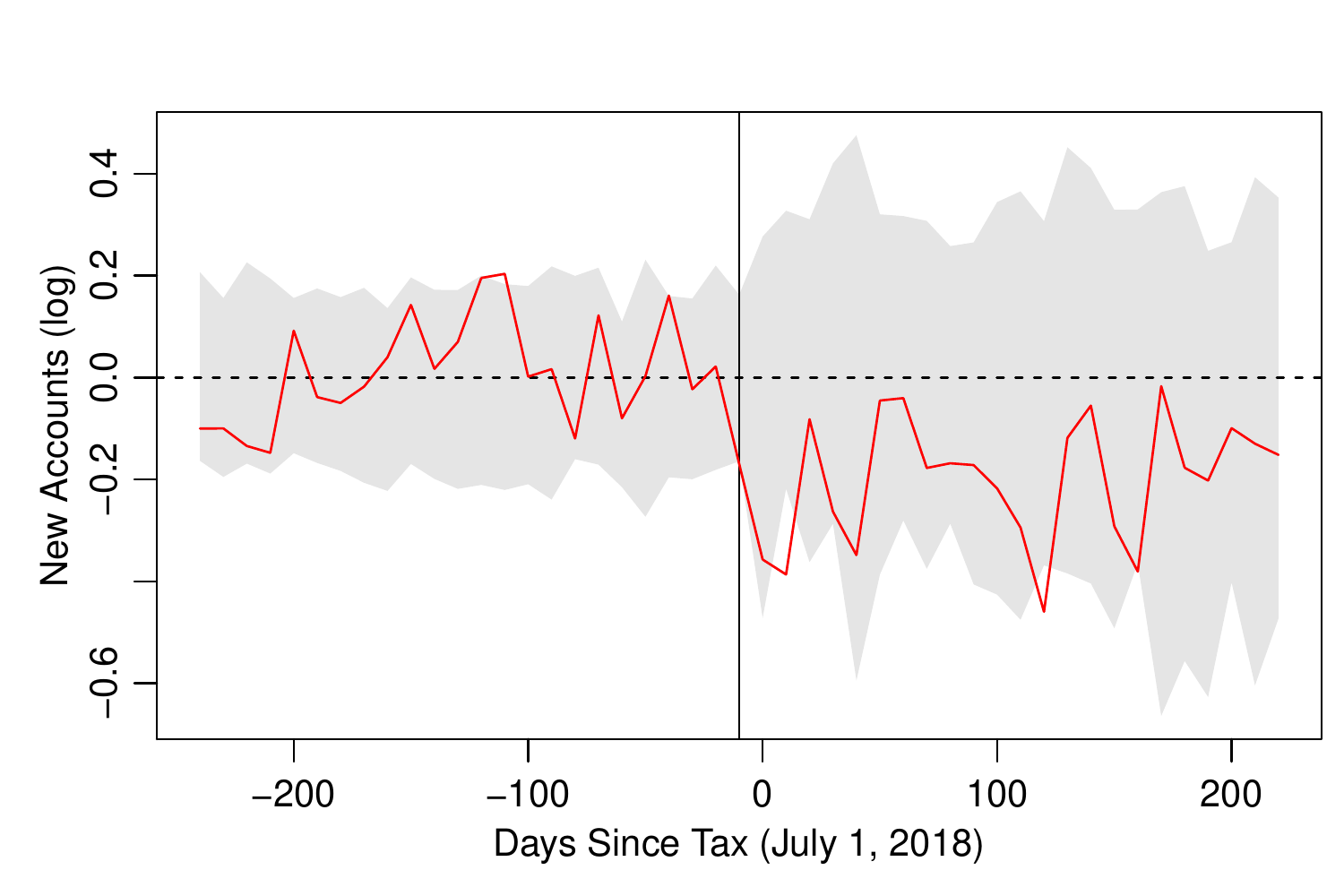}

\includegraphics[scale=0.33]{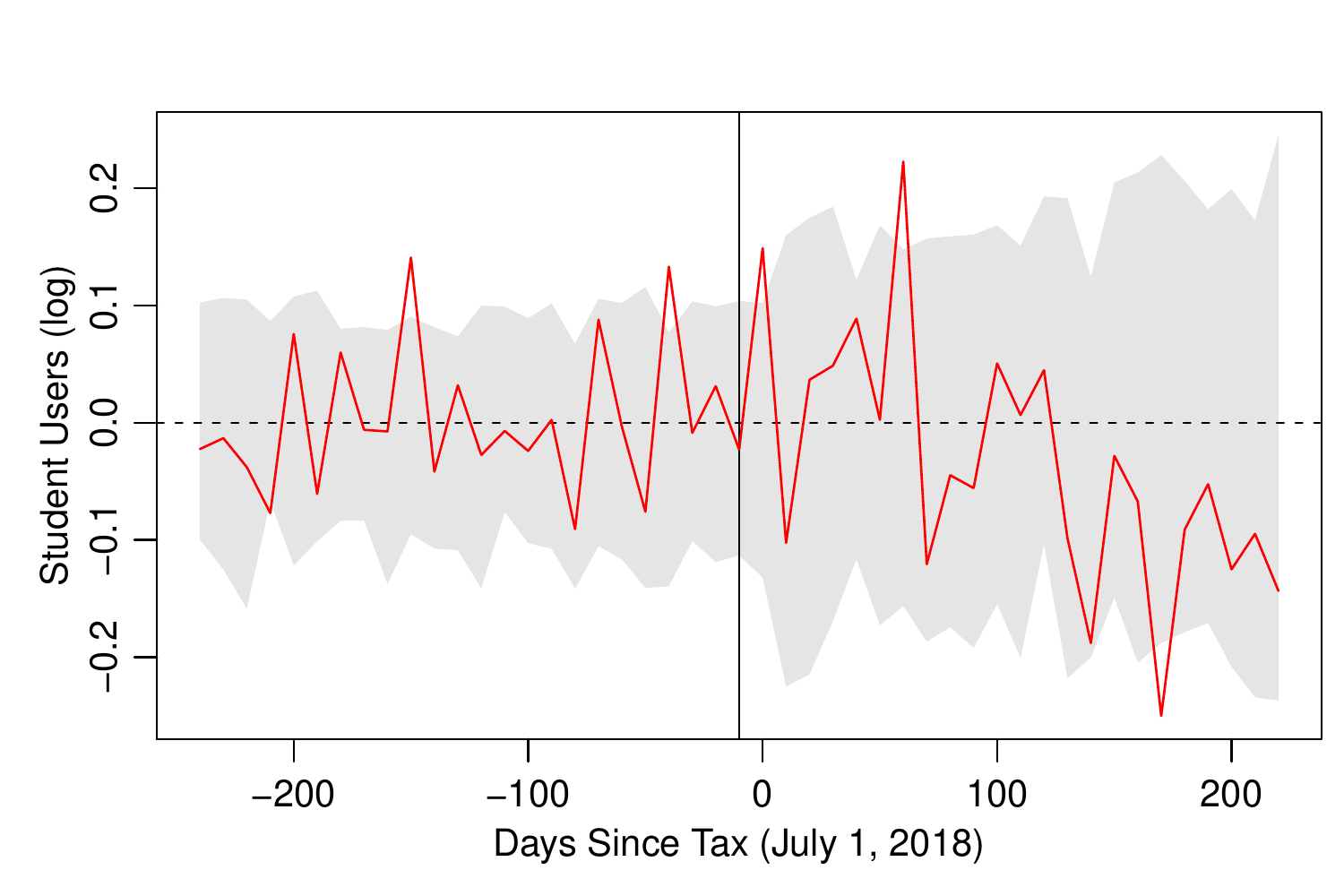}\includegraphics[scale=0.33]{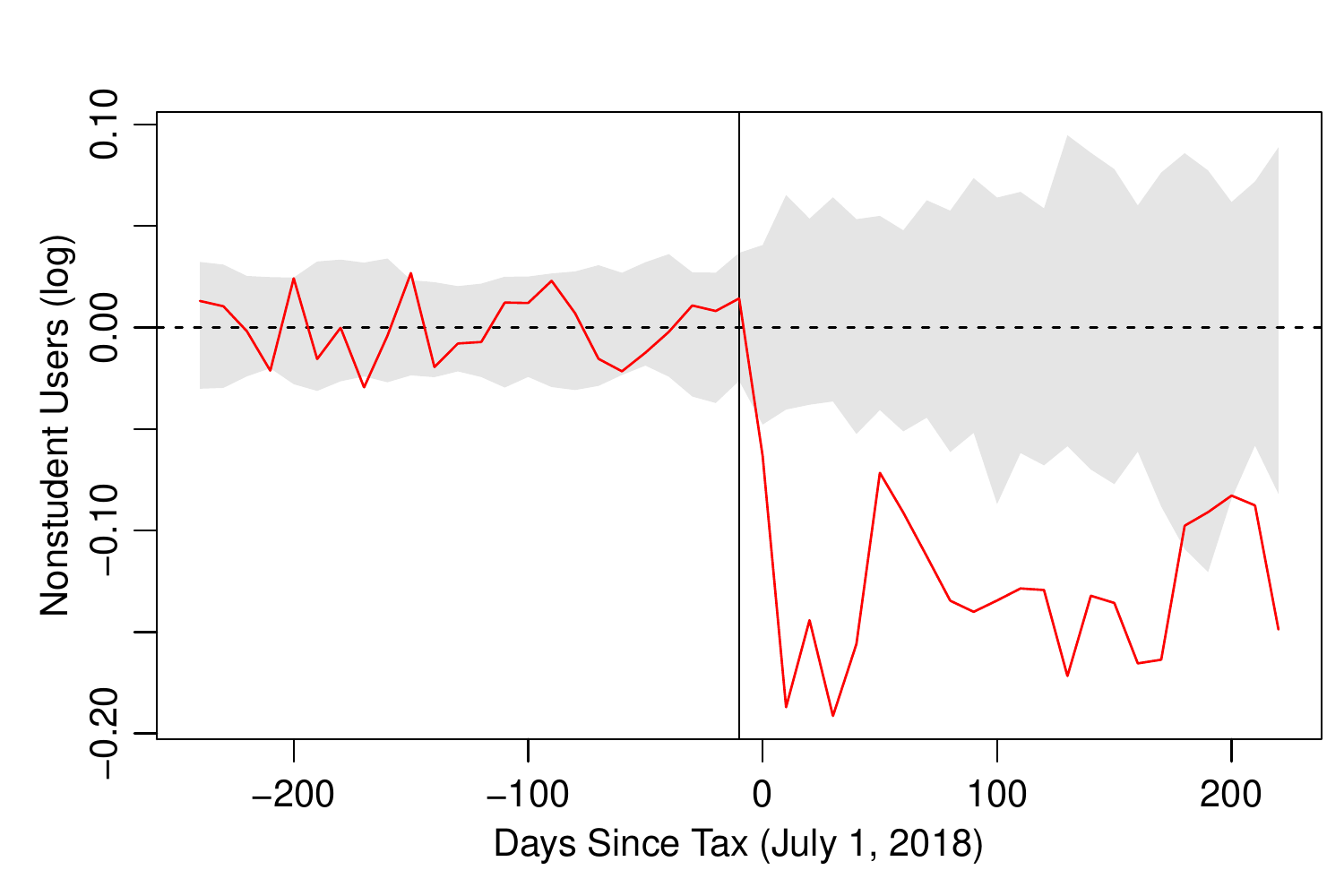}\includegraphics[scale=0.33]{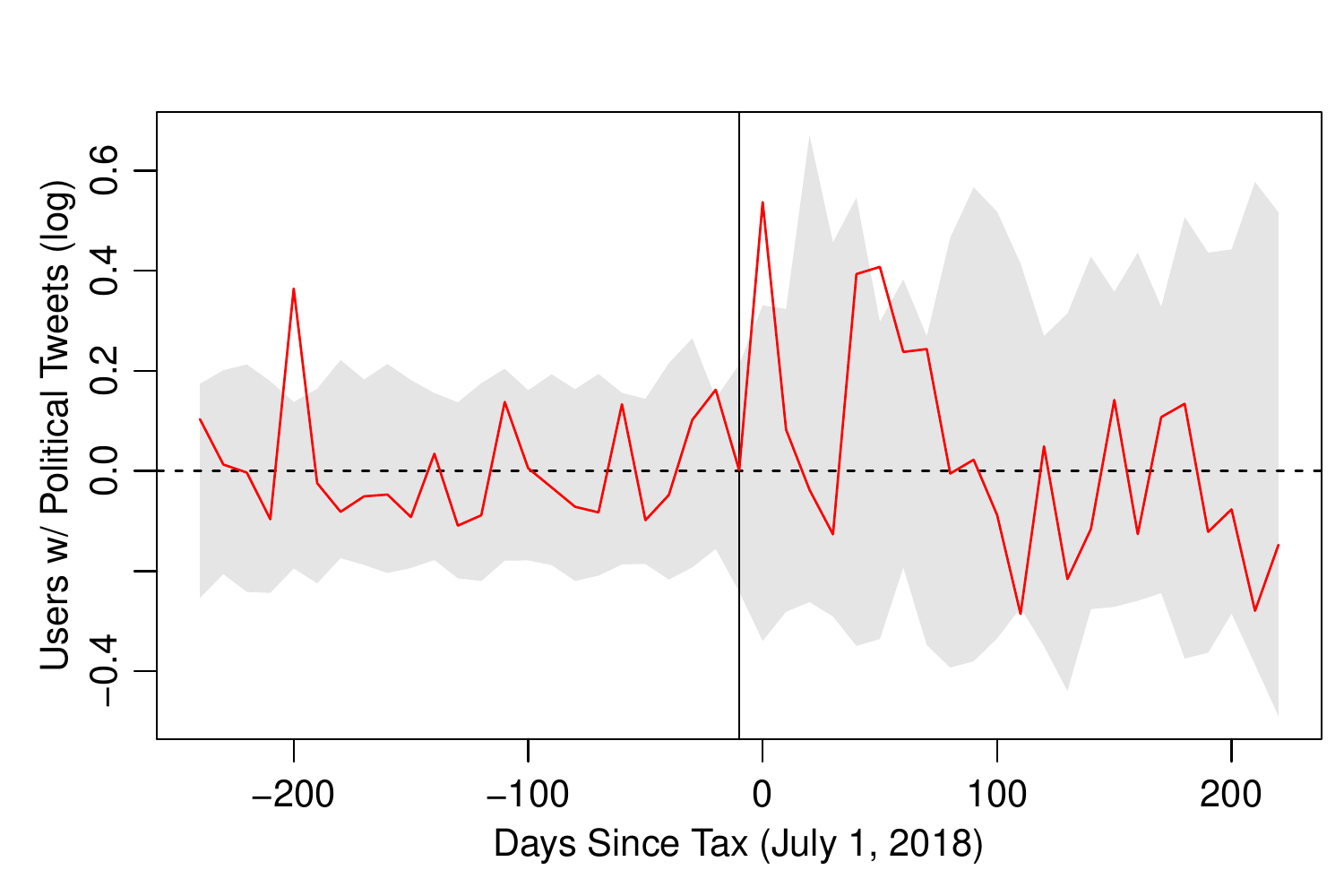}

\includegraphics[scale=0.33]{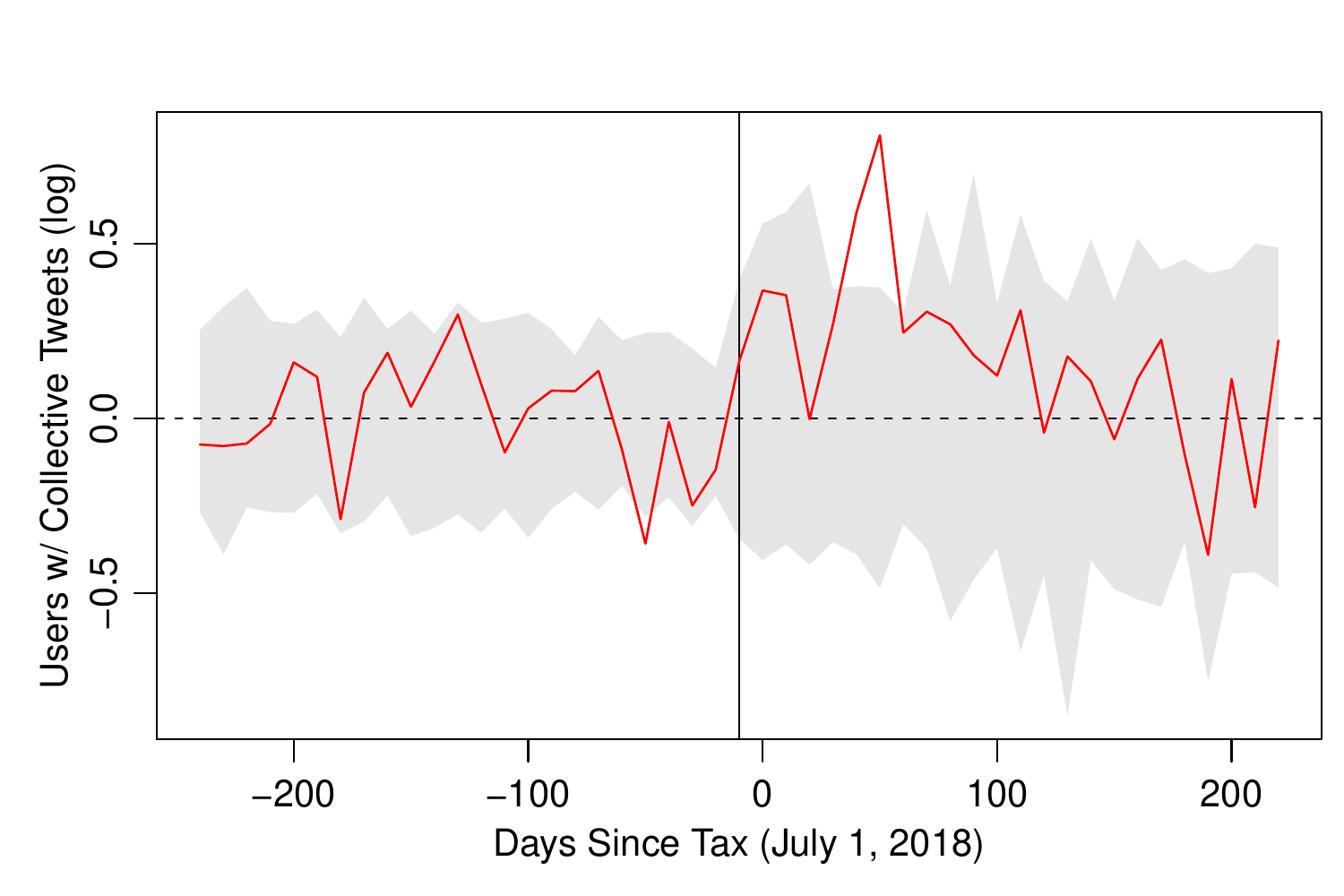}\includegraphics[scale=0.33]{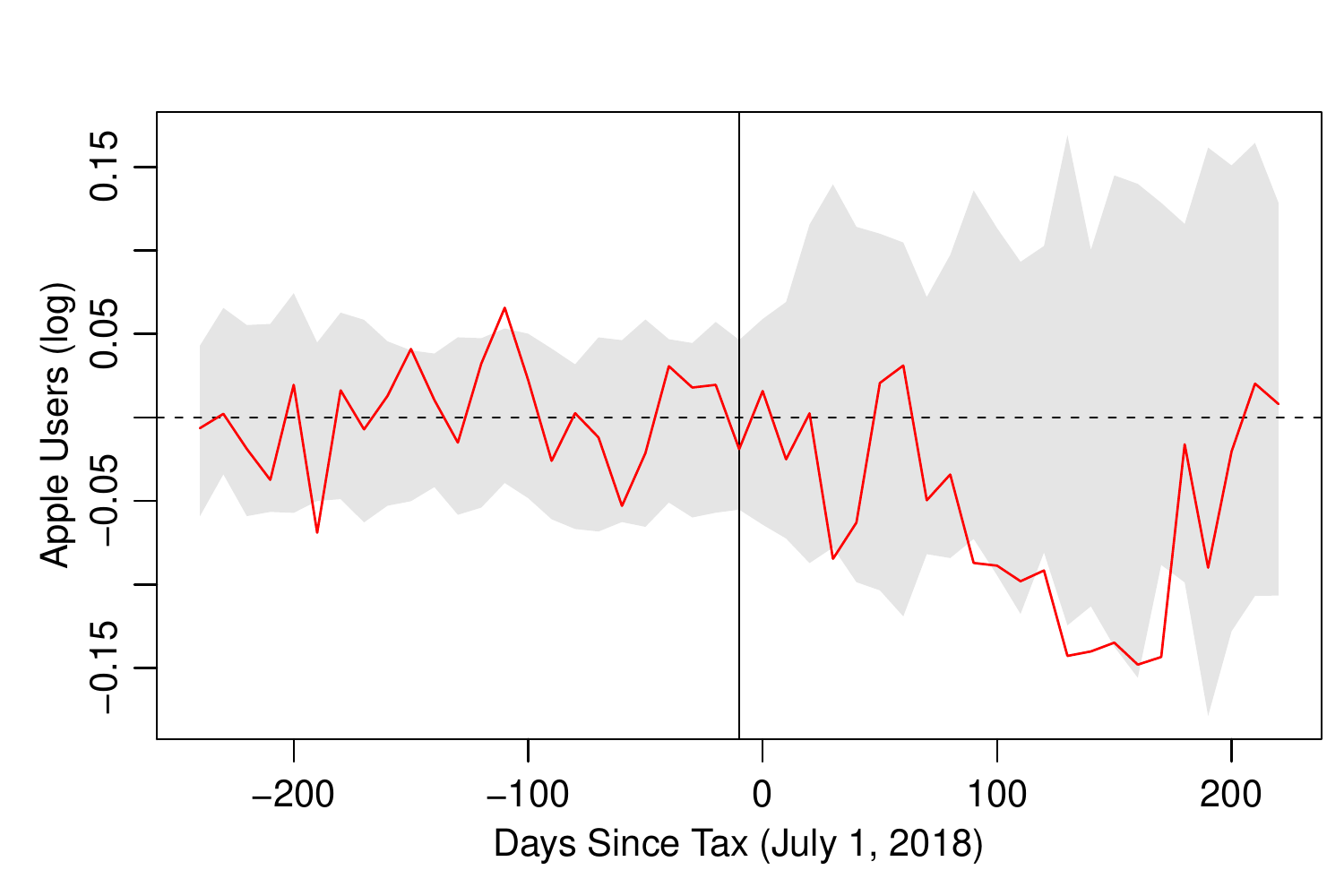}\includegraphics[scale=0.33]{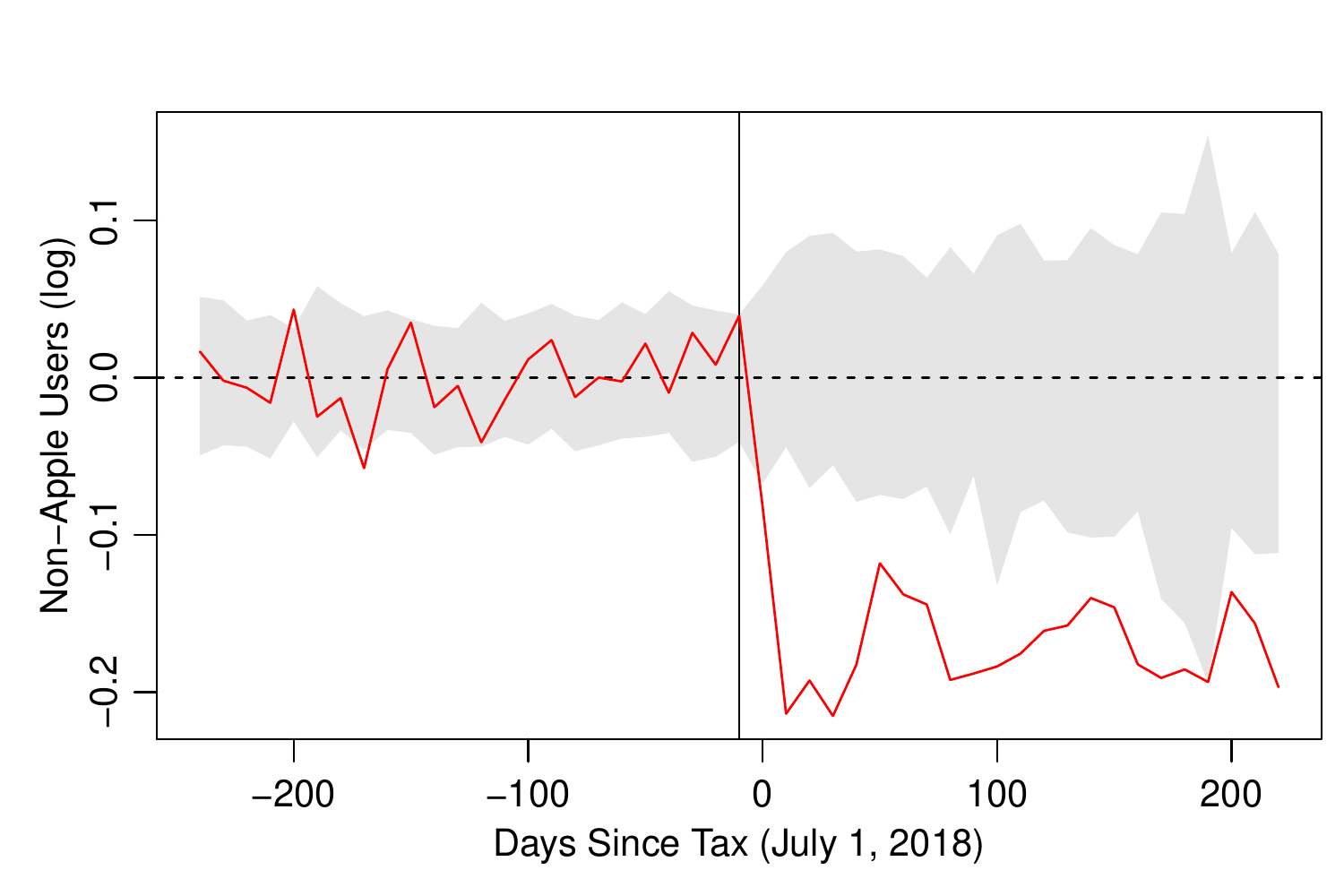}

\includegraphics[scale=0.33]{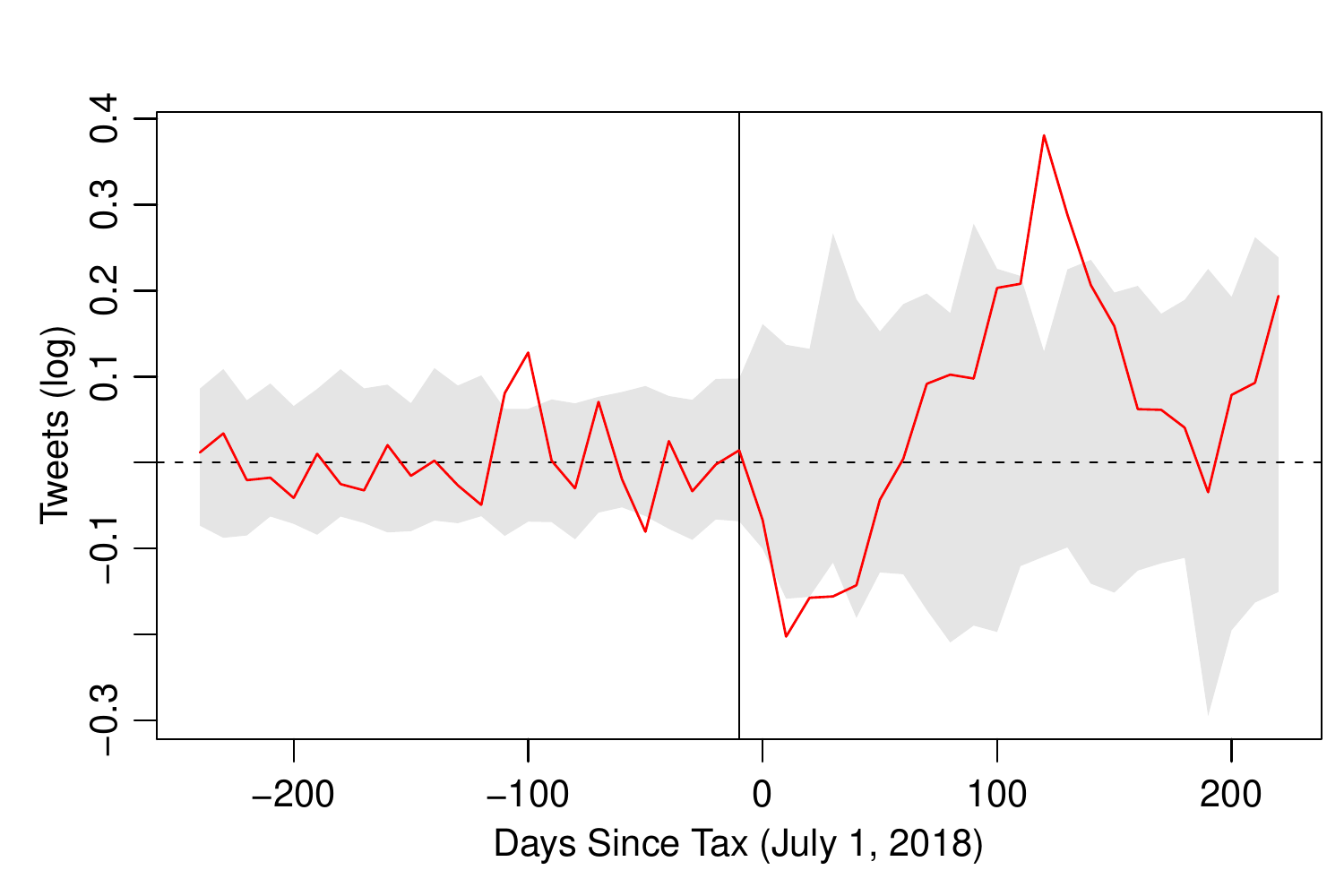}\includegraphics[scale=0.33]{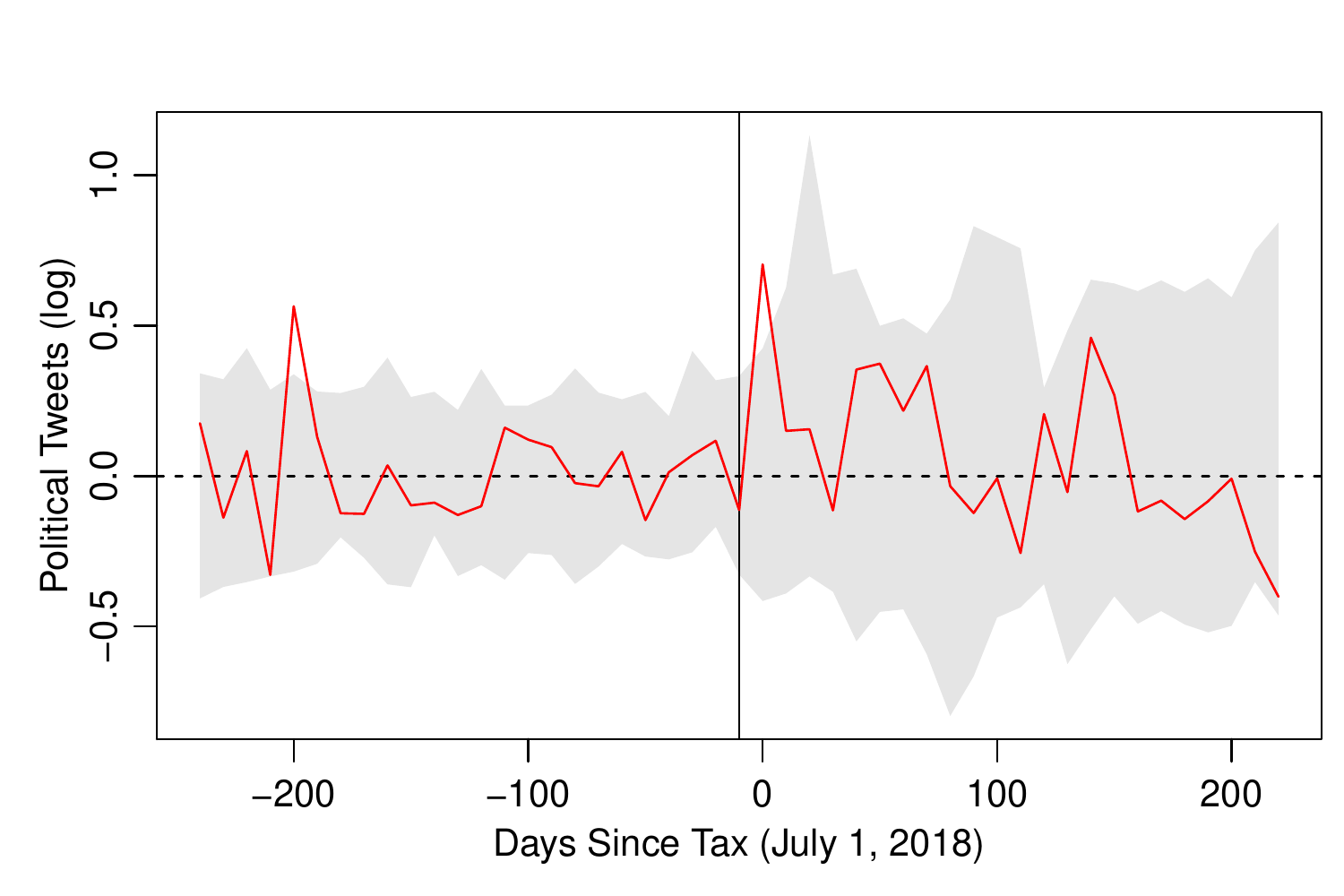}\includegraphics[scale=0.33]{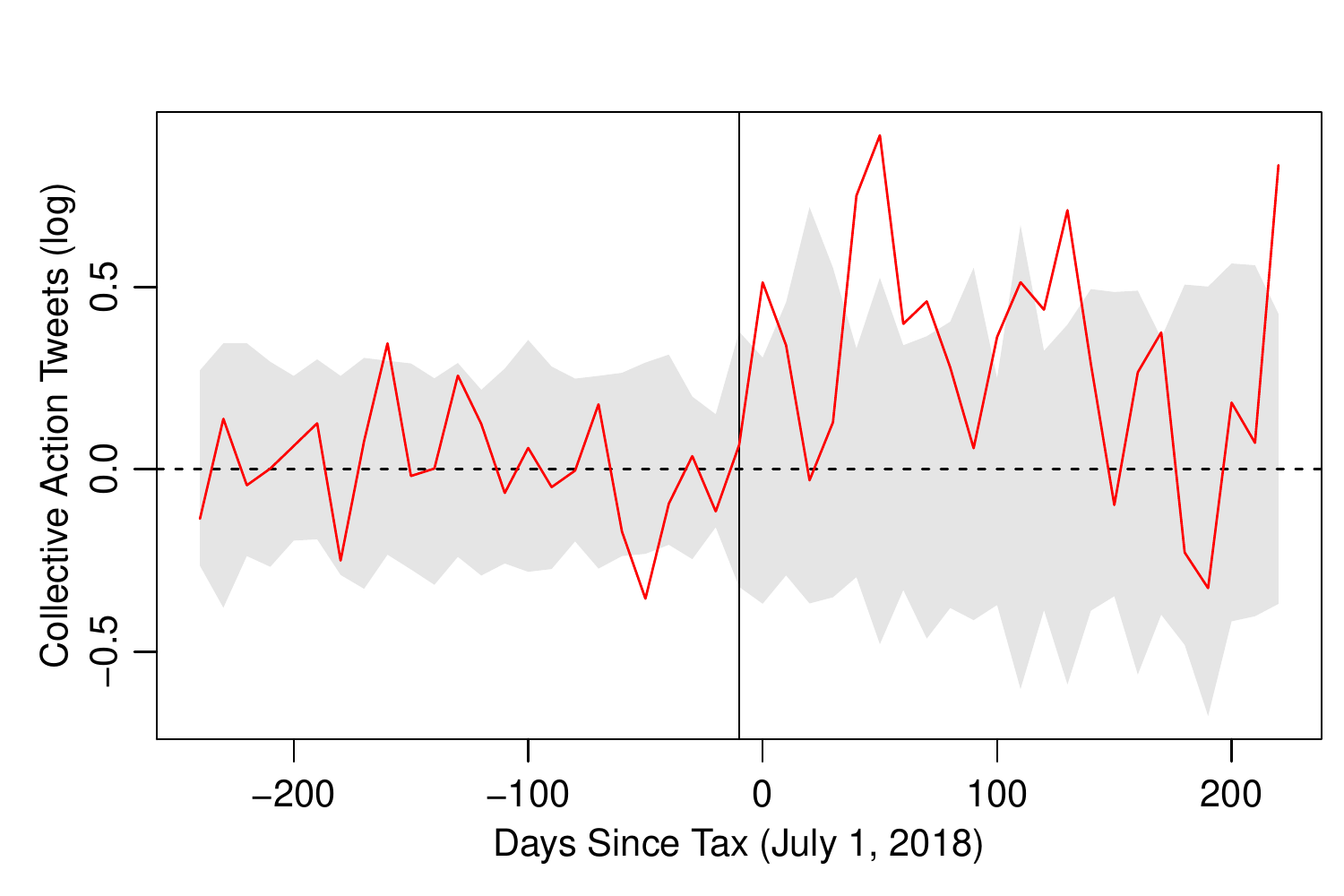}

\includegraphics[scale=0.33]{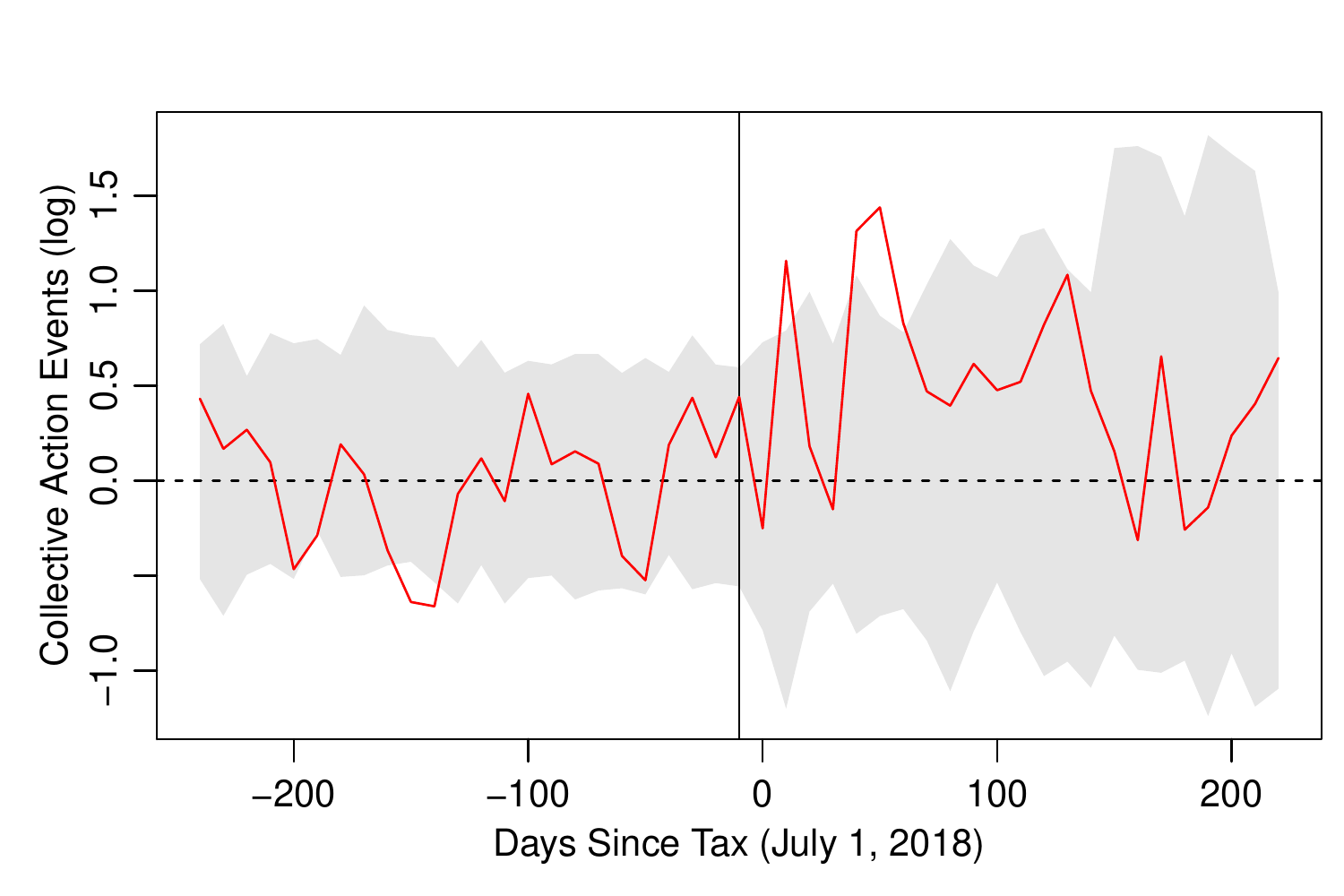}

{\footnotesize{}Notes: The figure reports the estimated treatment
effects of the social media tax on various outcomes in Uganda during
each ten day period relative to a synthetic control. The shaded region
indicates the .025 and .975 quantiles of the scaled placebo distribution
$\mathcal{P}_{t}$. The vertical line indicates the 10 day period
immediately prior to when the social media tax was implemented in
Uganda (July 1, 2018). See Section \ref{subsec:Additional-Data-Details}
for details on the construction of each outcome variable.}{\footnotesize\par}
\end{figure}

\begin{figure}[H]
\caption{Effect of the Tax on Collective Action \textendash{} Falsification
Test\label{fig:collective_action_plac}}

\medskip{}

\begin{centering}
\includegraphics{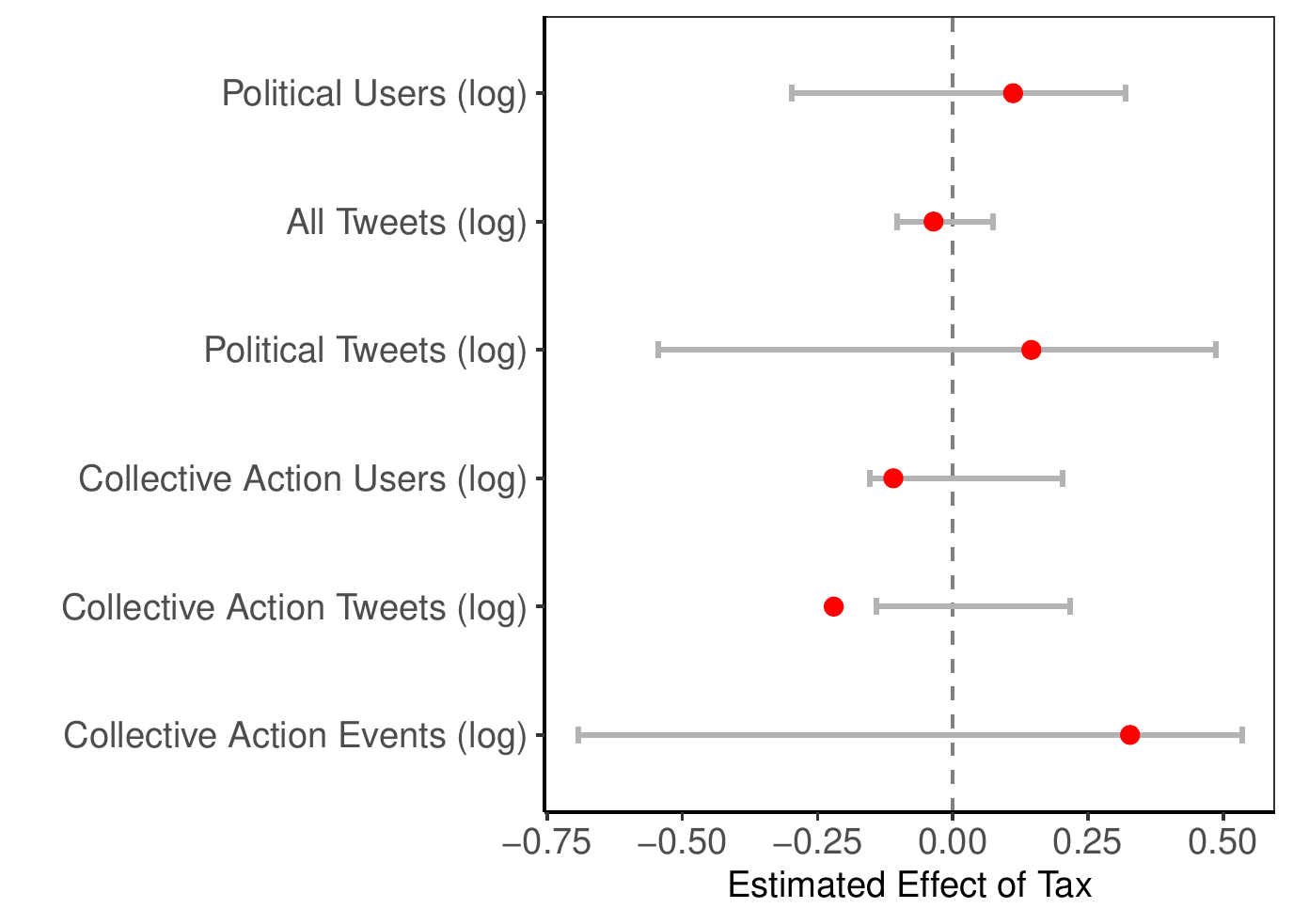}
\par\end{centering}
{\footnotesize{}Notes: The figure plots the average estimated treatment
effect (dot) of the social media tax on various outcomes along with
the .025 and .975 quantiles of the scaled placebo distribution $\mathcal{P}$
(grey bars) when restricting the model fit of the synthetic control
to more than 100 days prior to the tax. The average of the estimated
treatment effects are then taken over the 100 days prior to the tax.
See Section \ref{subsec:Additional-Data-Details} for further details
on each outcome and Figure \ref{fig:collective_action} to compare
to the true estimated treatment effects.}{\footnotesize\par}
\end{figure}

\begin{figure}[H]
\caption{Effect of the Tax on Collective Action \textendash{} Robustness\label{fig:collective_action_robust}}

\medskip{}

\begin{centering}
\includegraphics{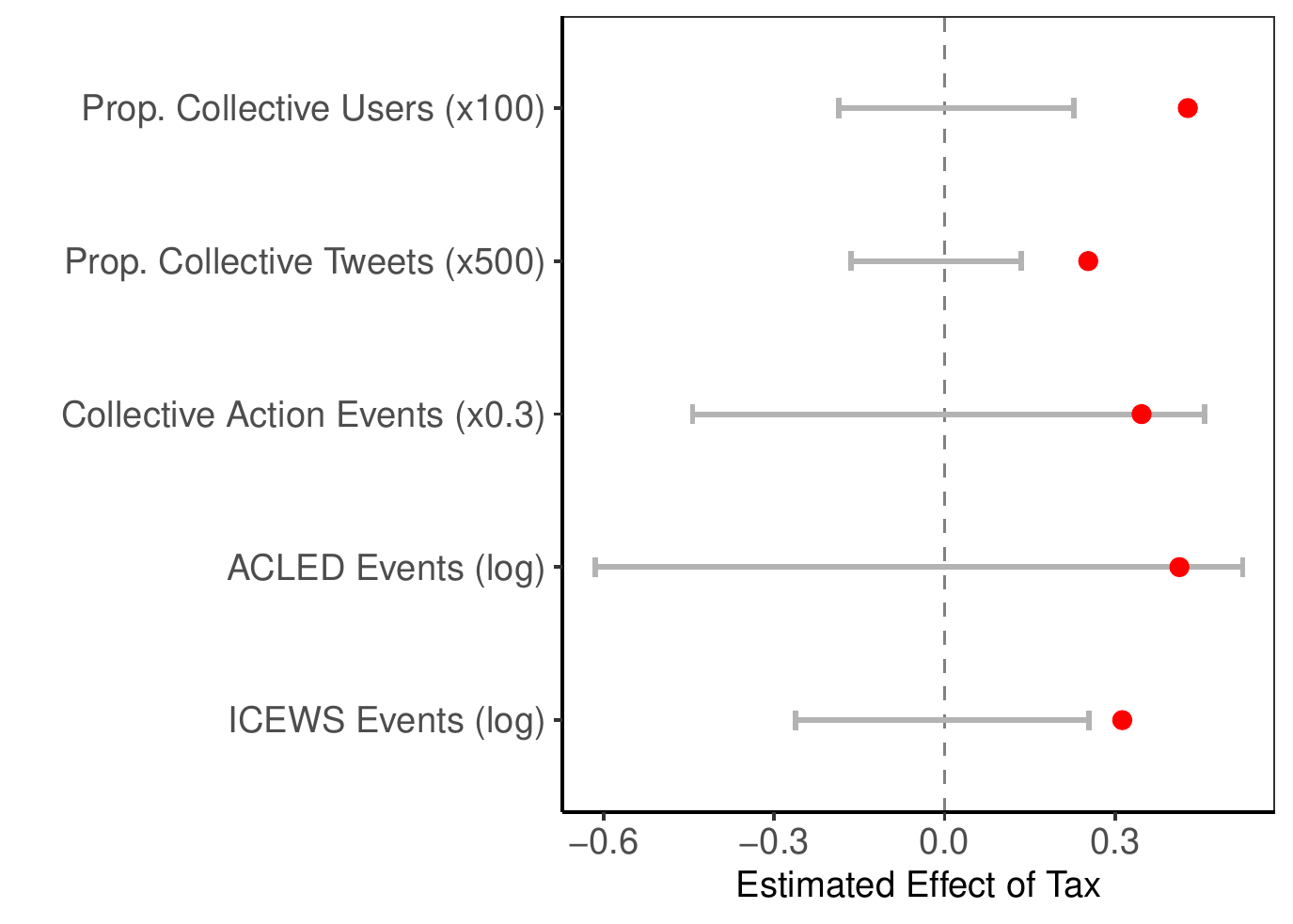}
\par\end{centering}
{\footnotesize{}Notes: The figure plots the average estimated treatment
effect }$\frac{1}{T_{1}+1}\sum_{t\geq0}\hat{\tau}_{0t}${\footnotesize{}
(dot) of the social media tax on various outcomes along with the .025
and .975 quantiles of the scaled placebo distribution $\mathcal{P}$
(grey bars). The `Prop. Collective Users,' `Prop. Collective Tweets,'
and `Collective Action Events' outcomes are scaled by 100, 500, and
0.3 respectively. The `Prop.' outcomes report the proportion of users
that are collective action users or the proportion of tweets that
are collective action tweets. See the main text and Section \ref{subsec:Additional-Data-Details}
for additional details on each outcome.}{\footnotesize\par}
\end{figure}

\pagebreak{}

\begin{figure}[H]
\caption{Collective Action Tweets Mentioning the Tax\label{fig:tax_and_collective}}

\medskip{}

\begin{centering}
\includegraphics[scale=0.75]{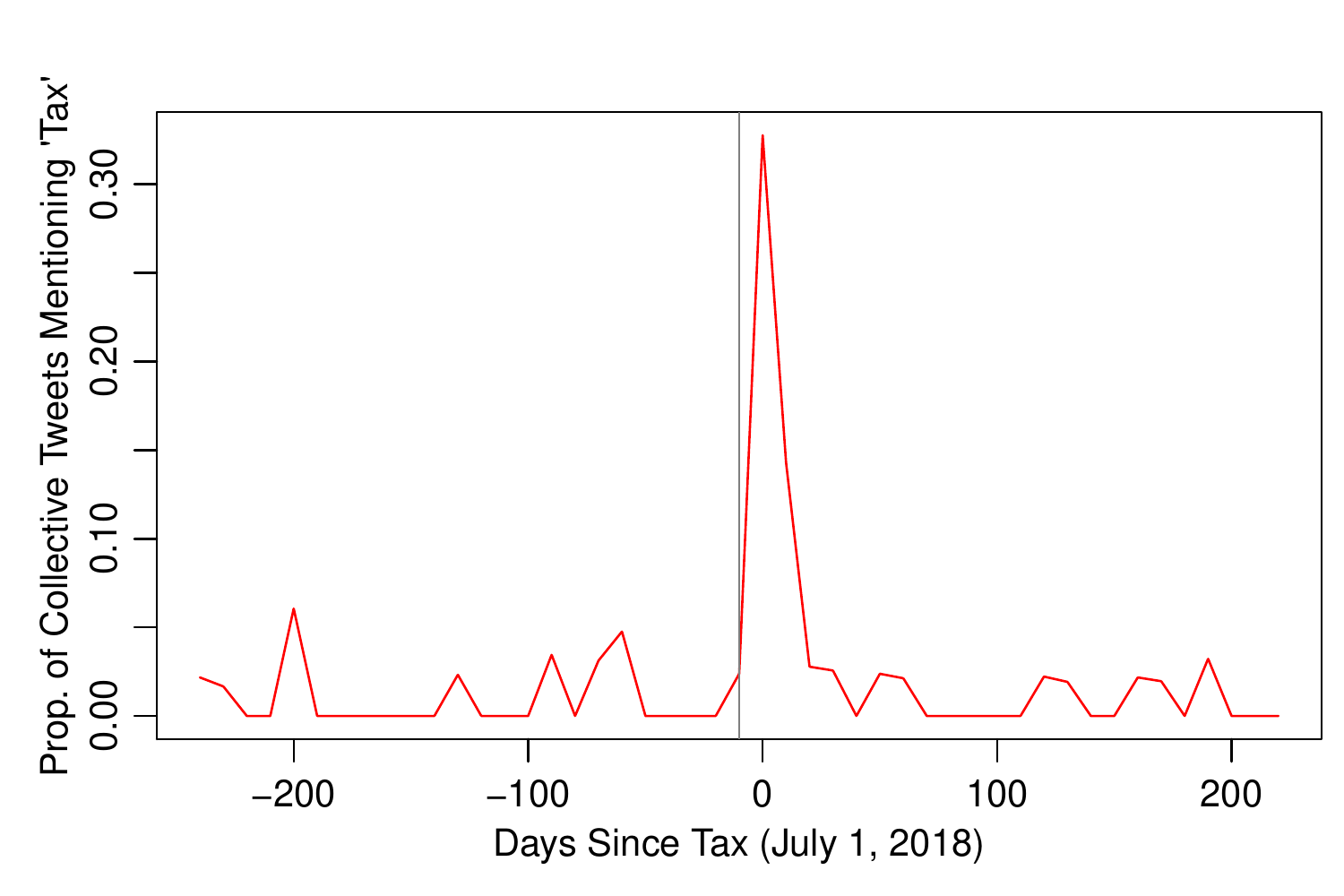}
\par\end{centering}
{\footnotesize{}Notes: The plot shows the proportion of collective
action tweets mentioning `tax' in Uganda in each 10 day period. The
solid vertical line indicates the 10 day period immediately prior
to when the social media tax was implemented in Uganda (July 1, 2018).}{\footnotesize\par}
\end{figure}

\begin{landscape}
\begin{center}
{\tiny{}}
\begin{table}[H]
{\tiny{}\caption{Top 10 Weights for Synthetic Control\label{tab:weights}}
}{\tiny\par}
\begin{centering}
{\small{}}%
\begin{tabular}{lccccccccccccc}
\hline 
\multicolumn{2}{c}{{\small{}Users (log)}} &  & \multicolumn{2}{c}{{\small{}Infrequent Users (log)}} &  & \multicolumn{2}{c}{{\small{}Activist Users (log)}} &  & \multicolumn{2}{c}{{\small{}Tweets (log)}} &  & \multicolumn{2}{c}{{\small{}Collective Action Tweets (log)}}\tabularnewline
\cline{1-2} \cline{2-2} \cline{4-5} \cline{5-5} \cline{7-8} \cline{8-8} \cline{10-11} \cline{11-11} \cline{13-14} \cline{14-14} 
{\small{}Country} & {\small{}Weight} &  & {\small{}Country} & {\small{}Weight} &  & {\small{}Country} & {\small{}Weight} &  & {\small{}Country} & {\small{}Weight} &  & {\small{}Country} & {\small{}Weight}\tabularnewline
\cline{1-2} \cline{2-2} \cline{4-5} \cline{5-5} \cline{7-8} \cline{8-8} \cline{10-11} \cline{11-11} \cline{13-14} \cline{14-14} 
{\small{}GH} & {\small{}0.312} &  & {\small{}GH} & {\small{}0.376} &  & {\small{}KE} & {\small{}0.302} &  & {\small{}KE} & {\small{}0.394} &  & {\small{}KE} & {\small{}0.315}\tabularnewline
{\small{}KE} & {\small{}0.288} &  & {\small{}KE} & {\small{}0.204} &  & {\small{}GH} & {\small{}0.214} &  & {\small{}CM} & {\small{}0.169} &  & {\small{}ZW} & {\small{}0.215}\tabularnewline
{\small{}RW} & {\small{}0.152} &  & {\small{}CD} & {\small{}0.128} &  & {\small{}ZW} & {\small{}0.143} &  & {\small{}TZ} & {\small{}0.112} &  & {\small{}ZM} & {\small{}0.210}\tabularnewline
{\small{}ET} & {\small{}0.059} &  & {\small{}ZM} & {\small{}0.113} &  & {\small{}AO} & {\small{}0.100} &  & {\small{}SN} & {\small{}0.064} &  & {\small{}SL} & {\small{}0.078}\tabularnewline
{\small{}SO} & {\small{}0.058} &  & {\small{}RW} & {\small{}0.105} &  & {\small{}ZM} & {\small{}0.064} &  & {\small{}GN} & {\small{}0.052} &  & {\small{}MU} & {\small{}0.052}\tabularnewline
{\small{}CD} & {\small{}0.049} &  & {\small{}ET} & {\small{}0.046} &  & {\small{}MZ} & {\small{}0.057} &  & {\small{}BJ} & {\small{}0.042} &  & {\small{}AO} & {\small{}0.050}\tabularnewline
{\small{}ZM} & {\small{}0.032} &  & {\small{}SO} & {\small{}0.029} &  & {\small{}RE} & {\small{}0.034} &  & {\small{}GH} & {\small{}0.035} &  & {\small{}SZ} & {\small{}0.025}\tabularnewline
{\small{}BI} & {\small{}0.024} &  & {\small{}TD} & {\small{}0.000} &  & {\small{}MU} & {\small{}0.032} &  & {\small{}BI} & {\small{}0.034} &  & {\small{}RW} & {\small{}0.021}\tabularnewline
{\small{}NE} & {\small{}0.023} &  & {\small{}SN} & {\small{}0.000} &  & {\small{}CV} & {\small{}0.028} &  & {\small{}TD} & {\small{}0.026} &  & {\small{}MW} & {\small{}0.020}\tabularnewline
{\small{}SN} & {\small{}0.002} &  & {\small{}ZW} & {\small{}0.000} &  & {\small{}SZ} & {\small{}0.026} &  & {\small{}ZW} & {\small{}0.018} &  & {\small{}BJ} & {\small{}0.008}\tabularnewline
\hline 
\end{tabular}{\small\par}
\par\end{centering}
{\small{}Notes: The table shows the top ten countries along with their
associated weights use for constructing the synthetic control for
each outcome variable. Country is the two-letter ISO 3166 country
code.}{\small\par}

\end{table}
{\tiny\par}
\par\end{center}

\end{landscape}
\begin{center}
{\tiny{}}
\begin{table}[H]
{\tiny{}\caption{Determinants of Collective Action and Political Tweets\label{tab:who_protests}}
}{\tiny\par}
\begin{centering}
{\small{}}%
\begin{tabular}{lcccccc}
\hline 
{\small{}Dependent Variable:} & \multicolumn{3}{c}{{\small{}Collective Action Tweets}} & \multicolumn{3}{c}{{\small{}Political Tweets}}\tabularnewline
\hline 
 & {\small{}(1)} & {\small{}(2)} & {\small{}(3)} & {\small{}(4)} & {\small{}(5)} & {\small{}(6)}\tabularnewline
\hline 
{\small{}Apple User $\times$ .01} & {\small{}0.077} & {\small{}25.804} & {\small{}0.001} & {\small{}-0.316} & {\small{}-10.660} & {\small{}-0.000}\tabularnewline
 & {\small{}(0.039)} & {\small{}(12.148)} & {\small{}(0.000)} & {\small{}(0.249)} & {\small{}(8.375)} & {\small{}(0.002)}\tabularnewline
{\small{}Frequent User $\times$ .01} & {\small{}0.017} & {\small{}5.985} & {\small{}0.038} & {\small{}-0.238} & {\small{}-7.310} & {\small{}-0.058}\tabularnewline
 & {\small{}(0.032)} & {\small{}(11.025)} & {\small{}(0.035)} & {\small{}(0.264)} & {\small{}(7.533)} & {\small{}(0.164)}\tabularnewline
{\small{}Student User $\times$ .01} & {\small{}-0.065} & {\small{}-25.099} & {\small{}-0.001} & {\small{}0.918} & {\small{}24.618} & {\small{}0.012}\tabularnewline
 & {\small{}(0.047)} & {\small{}(19.076)} & {\small{}(0.001)} & {\small{}(0.419)} & {\small{}(10.660)} & {\small{}(0.004)}\tabularnewline
{\small{}Longtime User $\times$ .01} & {\small{}0.106} & {\small{}38.093} & {\small{}0.116} & {\small{}0.372} & {\small{}10.607} & {\small{}0.001}\tabularnewline
 & {\small{}(0.041)} & {\small{}(15.834)} & {\small{}(0.035)} & {\small{}(0.333)} & {\small{}(9.650)} & {\small{}(0.173)}\tabularnewline
\hline 
{\small{}Level of Observation} & {\small{}Tweet} & {\small{}Tweet} & {\small{}10-day Period} & {\small{}Tweet} & {\small{}Tweet} & {\small{}10-day Period}\tabularnewline
{\small{}Estimator} & {\small{}OLS} & {\small{}Logit} & {\small{}OLS} & {\small{}OLS} & {\small{}Logit} & {\small{}OLS}\tabularnewline
{\small{}Period FE} & {\small{}Y} & {\small{}Y} & {\small{}Y} & {\small{}Y} & {\small{}Y} & {\small{}Y}\tabularnewline
{\small{}Location FE} & {\small{}Y} & {\small{}Y} & {\small{}Y} & {\small{}Y} & {\small{}Y} & {\small{}Y}\tabularnewline
{\small{}Clusters} & {\small{}17016} & {\small{}16646} & {\small{}17016} & {\small{}17016} & {\small{}17007} & {\small{}17016}\tabularnewline
{\small{}Observations} & {\small{}483090} & {\small{}476652} & {\small{}81393} & {\small{}483090} & {\small{}482860} & {\small{}81393}\tabularnewline
\hline 
\end{tabular}{\small\par}
\par\end{centering}
{\small{}Notes: Table shows the results of regressing measures of
collective action or political tweets on user and period characteristics.
Columns (1) and (2) use an indicator for whether a tweet contains
a collective action phrase. Column (3) collapses observations by user
id, location, and 10-day periods by taking the mean of each variable
over the associated set of tweets. Columns (4)\textendash (6) are
analogous, but use a measure of whether a tweet contains a political
phrase not related to collective action. Fixed effects for 10-day
periods and location name as assigned by Twitter are included throughout.
Observations are restricted to the period prior to the tax, and to
tweets where both the user language and tweet language is specified
as English. All independent variables reported are scaled by .01.
`Apple User' is an indicator for whether the tweet came from an Apple
device; for Columns (3) and (6), it is an indicator for whether a
user tweeted from an Apple device at least once during the period.
`Frequent User' is an indicator for whether an individual tweeted
more than once per day between joining Twitter and when they initially
appear in our data. `Student User' is an indicator for whether the
user included a university-related phrase in their description or
location at the time of the tweet; for Columns (3) and (6), it is
an indicator for whether a university-related phrase appears in the
description or location for at least one tweet during the period.
`Longtime User' is an indicator for whether an individual joined Twitter
more than four years prior to the start of the tax. Coefficients are
reported throughout, and standard errors reported in parentheses below
are clustered at the user level.}{\small\par}
\end{table}
{\tiny\par}
\par\end{center}

\end{document}